\documentclass[a4paper,fleqn,usenatbib,useAMS]{mnras}

\usepackage{graphicx}
\usepackage{amssymb}
\usepackage{subfigure}
\usepackage{captcont}
\usepackage{multirow}
\usepackage{longtable}
\usepackage{comment}
\usepackage{array}
\usepackage{url}
\usepackage{pdflscape}
\usepackage{booktabs,multirow}
\usepackage{times} 
\usepackage{units}
\usepackage{aas_macros} 

\usepackage[T1]{fontenc}
\usepackage{aecompl}
\usepackage{newtxtext,newtxmath}

\usepackage[usenames]{color}

\usepackage{upgreek}

\newcommand{\cd}{d$^{-1}$}

\title[Phase suppression in the roAp star J1640]{Suppressed phase variations in a high amplitude rapidly oscillating Ap star pulsating in a distorted quadrupole mode}
\author[D. L. Holdsworth et al.]{Daniel L. Holdsworth,$^{1,2}$\thanks{E-mail: dlholdsworth@uclan.ac.uk}
H. Saio,$^{3}$
D. M. Bowman,$^{4}$
D. W. Kurtz,$^{1}$
R. R. Sefako,$^{5}$
\newauthor
M. Joyce,$^{6,5}$
T. Lambert$^{7}$
and B. Smalley$^{8}$\\
$^{1}$ Jeremiah Horrocks Institute, University of Central Lancashire, Preston PR1 2HE, UK\\
$^{2}$ Department of Physics, North-West University, Mafikeng Campus, Private Bag X2046, Mmabatho 2745, South Africa\\
$^{3}$ Astronomical Institute, School of Science, Tohoku University, Sendai 980-8578, Japan\\
$^{4}$ Instituut voor Sterrenkunde, KU Leuven, Celestijnenlaan 200D, B-3001 Leuven, Belgium\\
$^{5}$ South African Astronomical Observatory, PO Box 9, Observatory, Cape Town 7935, South Africa\\
$^{6}$ Department of Physics and Astronomy, Dartmouth College, Hanover, NH 03755, USA\\
$^{7}$ Department of Astronomy, University of Cape Town, 7700 Rondebosch, South Africa\\
$^{8}$ Astrophysics Group, Keele University, Staffordshire ST5 5BG, UK
}
\begin{document}

\date{\today}

\pagerange{\pageref{firstpage}--\pageref{lastpage}} \pubyear{2018} 

\maketitle

\label{firstpage}

\begin{abstract}
We present the results of a multisite photometric observing campaign on the rapidly oscillating Ap (roAp) star 2MASS\,16400299-0737293 (J1640; $V=12.7$). We analyse photometric $B$ data to show the star pulsates at a frequency of $151.93$\,\cd\, ($1758.45\,\muup$Hz; $P=9.5$\,min) with a peak-to-peak amplitude of 20.68\,mmag, making it one of the highest amplitude roAp stars. No further pulsation modes are detected. The stellar rotation period is measured at $3.6747\pm0.0005$\,d, and we show that rotational modulation due to spots is in anti-phase between broadband and $B$ observations. Analysis and modelling of the pulsation reveals this star to be pulsating in a distorted quadrupole mode, but with a strong spherically symmetric component. The pulsational phase variation in this star is suppressed, leading to the conclusion that the contribution of $\ell>2$ components dictate the shape of phase variations in roAp stars that pulsate in quadrupole modes. This is only the fourth time such a strong pulsation phase suppression has been observed, leading us to question the mechanisms at work in these stars. We classify J1640 as an A7\,Vp\,SrEu(Cr) star through analysis of classification resolution spectra.
\end{abstract}


\section{Introduction}
\label{sec:intro}

There exists a group of chemically peculiar A stars, known as Ap stars, that show overabundances of elements such as La, Pr, Eu, Sr, Cr and Nd in their atmospheres, when compared to their `normal' A-star counterparts; in some cases, these elements can be overabundant up to one million times the solar value \citep{ryabchikova04}. Often, these abundance anomalies form surface spots that modulate the observed flux of the star as it rotates. The peculiarities in the Ap stars arise from chemical stratification in the presence of a strong, global, magnetic field, which suppresses convection and allows radiative levitation. The magnetic field strengths in Ap stars are often in the range of a few kG, but can be as high as about 30\,kG (\citealt{babcock60}; \citealt{elkin2010}). It is common to find the magnetic axis misaligned with the rotation axis in Ap stars, leading to the oblique rotator model \citep{stibbs50}. In general, as a result of magnetic braking, the Ap stars are much slower rotators than their non- (or weakly) magnetic counterparts \citep{stepien00,abt95}; some of them are so strongly braked that their rotation periods can be decades, or even centuries \citep{mathys2015}.

A subset of the Ap stars are known to show short-period variability, on the order of 5--24\,min. These are the rapidly oscillating Ap (roAp) stars. Since their discovery by \citet{kurtz82}, only 61 roAp stars have been discussed in the literature \citep[for a catalogue see][]{smalley15,joshi16}. The driving mechanism of the pulsations in the roAp stars is thought to be the $\kappa$-mechanism acting in the H\,{\sc{i}} ionisation zone \citep[e.g.][]{balmforth01,saio05}, which results in high-overtone pressure modes (p\,modes) being excited. However, this mechanism cannot explain all the pulsations observed in roAp stars. Rather, \citet{cunha13} suggested that some observed modes may be excited by turbulent pressure in the convective zone. Furthermore, there is a subset of roAp stars that seem to pulsate with frequencies much higher than expected -- higher than the acoustic cutoff frequency. Work by e.g. \citet{kurtz94b,kurtz05,sachkov08,balona13,holdsworth16,holdsworth18a} has identified these stars, with some authors postulating whether there is yet another driving mechanism at work.

Since the first observations of the roAp stars \citep{kurtz82}, it was clear that a relation existed between the pulsation amplitude (and phase) and the rotation period of the star. This lead to the development of, and later enhancement of, the oblique pulsator model \citep{kurtz82,ss85a,ss85b,dg85,shibahashi93,ts94,ts95,bigot02,bigot11}. In this model, the pulsation axis is misaligned with the rotation one, but closely aligned to the magnetic axis. Such a configuration means that the viewing aspect of the pulsation mode changes over the rotation cycle of the star, leading to the observed amplitude and phase modulations. This property of the roAp stars can provide constraints on the geometry of observed pulsations.

Most roAp stars have been discovered through photometric campaigns targeting known Ap stars \citep[e.g.][]{martinez91,handler99,dorokhova05,paunzen15}, with some later additions through the use of high-resolution, time-resolved spectroscopy \citep[e.g.][]{hatzes04,elkin11,kochukhov13}. In more recent years, the use of ground-based surveys {\citep{holdsworth14a,holdsworth15} and the {\it Kepler} space telescope \citep[e.g.][]{kurtz11,balona11b,smalley15} have yielded new results. However, issues with using ground-based surveys are the sparsity and low-precision of the data. This method is useful for identifying pulsating Ap stars, but precise, high-cadence, follow-up observations are needed to fully study the targets.

The subject of this paper, J1640 ($\alpha$:\,16:40:02.99, $\delta$:\,$-07$:37:29.7; 2MASS\,J16400299-0737293; $V=12.7$), was first shown to be a roAp star by \citet{holdsworth14a} after searching the SuperWASP archive for periodic A star variables. Their results show variability at 151.93\,\cd\, ($1758.45\,\muup$Hz; $P=9.5$\,min) with a semi-amplitude of 3.52\,mmag in the SuperWASP broadband filter. With such a high amplitude pulsation, we consider whether J1640 is similar to the other roAp stars which exhibit similar pulsation amplitudes. Most pulsation amplitudes of roAp stars are below about 10\,mmag \citep{joshi16}, thus making the subset of high-amplitude pulsators interesting in their own right. Furthermore, these stars KIC\,7582608, HD\,24355 and J1940 \citep{holdsworth14b,holdsworth16,holdsworth18a}, are all single mode, distorted quadrupole pulsators which show suppressed pulsation phase variations, meaning that the pulsation cannot be modelled by a single spherical harmonic. These stars require energy from higher $\ell$ terms to accurately model the phase variations. As J1640 shows many similarities with these stars, we aim to test whether it also belongs to this abnormal group of roAp stars, and thus provide a larger subset of stars to investigate the pulsation driving mechanism in these peculiar class members. 

In the present paper, we provide a detailed analysis of the WASP discovery data, along with newly obtained, multisite, ground-based observations. Finally, we apply theoretical modelling to the data to better understand this pulsating star.


\section{Classification Spectroscopy}
\label{sec:spec}

To classify J1640, we have spectroscopically observed the star with two different instruments. The first spectrum was obtained with the Intermediate dispersion Spectrograph and Imaging System\footnote{\url{http://www.ing.iac.es/Astronomy/instruments/isis/}} (ISIS) mounted on the William Herschel Telescope (WHT). The second was obtained with the Robert Stobie Spectrograph \citep[RSS; ][]{kobulnicky03} mounted on the Southern African Large Telescope (SALT). We provide a log of the spectral observations in Table\,\ref{tab:spec}. 

\begin{table*}
  \caption{Log of spectroscopic observations of J1640. The {\sc{der\_snr}} code \citep{stoehr08} was used to find the S/N. The rotation phase has been calculated from equation\,(\ref{equ:rot}). BJD is the midpoint of the exposure, given as BJD-245\,0000.}
   \centering
  \label{tab:spec}
  \begin{tabular}{cccccc}
    \hline
    Instrument & BJD & Integration time & S/N & Resolution & Rotation\\
                       &       &   (s)         &     &  (\AA)          & phase   \\
    \hline
    WHT/ISIS &  6326.7699 & 1200 & 100 & 0.86 & $0.32\pm0.03$ \\
    SALT/RSS & 6374.5688 & 600 & 70 & 0.37  & $0.06\pm0.03$ \\
        \hline
    \end{tabular}
\end{table*}

We present the two spectra in Fig.\,\ref{fig:spec} along with three MK standard stars\footnote{Standard star spectra from R.O. Gray's website: \url{http://stellar.phys.appstate.edu/Standards/}} for spectral classification. We determine that the Balmer lines of J1640 are best represented by the A7 standard. However, the metal lines of J1640 do not match those of the standard star, as is expected in an Ap star. In particular, lines of Sr\,{\sc ii} at $4077$ and $4216$\,\AA\, are enhanced, Eu\,{\sc{ii}} at $4130$\,\AA\, is also stronger than the standard, as is the line at $4205$\,\AA. There is a weak signature of Cr\,{\sc ii} at 4111\,\AA, on the red side of the H$_\delta$ core. These deviations allow us to conclude that J1640 is an A7\,Vp\,SrEu(Cr) star. Additionally, we note a weak Mg\,{\sc ii} 4481\,\AA\, line when compared to the standard, as well as the commonly observed weak Ca\,{\sc{ii}} K line, a trait of Ap stars. Finally, there are slight line strength variations of some of the lines, most notably the Sr\,{\sc{ii}} lines, between the two spectra of J1640. This is a result of the difference in rotation phase at which the two spectra were obtained.

\begin{figure}
\includegraphics[width=\linewidth]{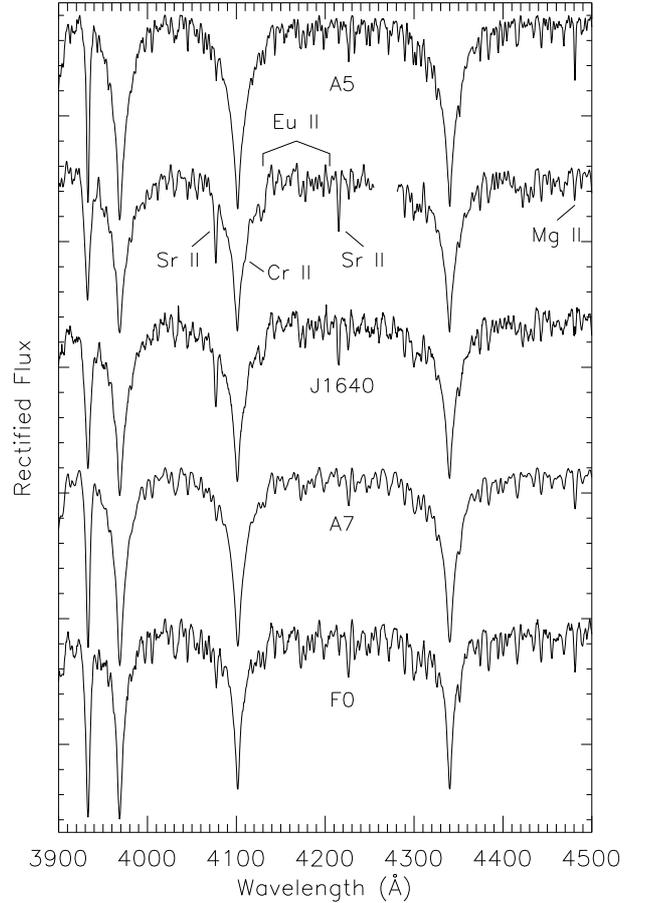}
\caption{Spectra of J1640 (second and third) compared to MK standard stars as labeled. The target spectra have been reduced to the resolution of the standards (i.e. 1.67\,\AA).}
\label{fig:spec}
\end{figure}

To estimate the temperature of J1640, we produce model spectra at a fixed $\log g=4.0$ (cgs) with the {\sc{uclsyn}} code \citep{smith88,smith92,uclsyn} and the VALD 2014 atomic line list, and try to best fit the Balmer lines. Our low-resolution data do not allow for the determination of $\log g$, thus we fix it at this reasonable value. We derive a temperature of $7600\pm200$\,K, and show the fit in Fig\,\ref{fig:spec_fit}. The cores of the hydrogen Balmer lines are not fitted well by the model. This is a known problem with LTE models which under-predict the depth of strong lines that are formed high in the atmosphere and subject to non-LTE effects \citep{gray08,smalley14c}. Coupled with this, the so-called core-wing anomaly is observed in Ap stars \citep{cowley01,kochukhov02}. In that case, however, the cores of hydrogen lines are deeper than models, which is not the case here. The underlying cause of the core-wing anomaly is elemental stratification, which leads to non-standard temperature gradients in Ap stars. Hence, in our fits we have ignored the cores of the Balmer lines. Also note in Fig.\,\ref{fig:spec_fit} how weak the Ca K line is with respect to the model. As mentioned earlier, this is a common trait of Ap stars.

\begin{figure}
\includegraphics[width=\linewidth]{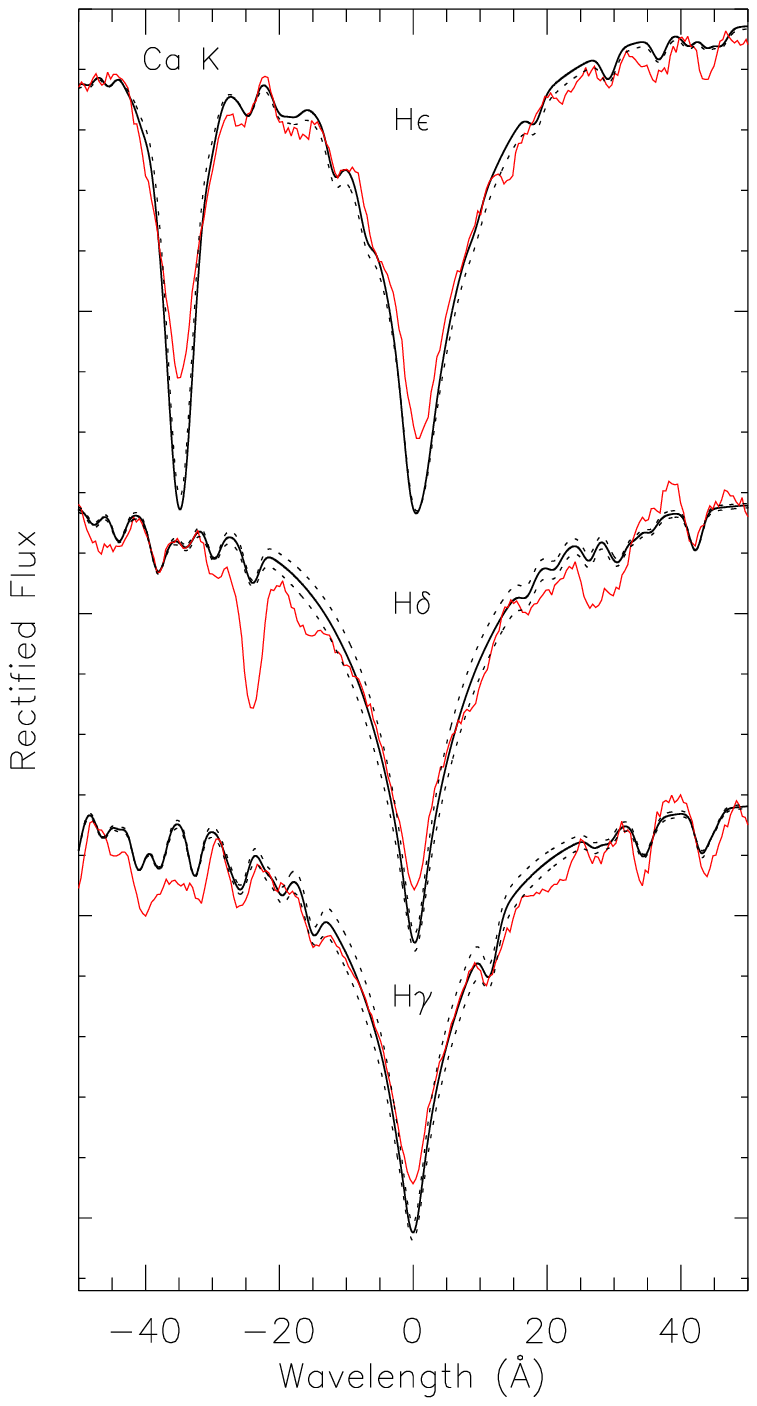}
\caption{The Balmer lines of J1640 (red) compared to a model spectrum of $T_{\rm eff} =7600$\,K, $\log g=4.0$ and solar metallicity (black solid line) to determine the effective temperature. Spectra with $\pm200$\,K are plotted with dotted lines to represent the error on the fit. Note the strong line of Sr\,{\sc{ii}} in the blue wing of H\,$_\delta$, and Eu\,{\sc{ii}} in the red wing. The wavelength is given relative to the core of each Hydrogen line.}
\label{fig:spec_fit}
\end{figure}

To determine the rotation phases at which the spectra were observed, we use the first rotational light maximum in the WASP data as the zero-point, and the rotation period derived in Section\,\ref{sec:SAAO_rot}, such that:
\begin{equation}
\phi(E) = (245\,5334.4160\pm0.0009) +(\mbox{3{\fd}6747}\pm\mbox{0{\fd}0005})\times E,
\label{equ:rot}
\end{equation}
where $E$ is the number of rotation cycles elapsed since the zero-point. The calculated rotation phases are shown in Table\,\ref{tab:spec}.

    
\section{Photometric observations}
\label{sec:phot}

We present the original SuperWASP discovery data here, from \citet{holdsworth14a}, with a more in-depth discussion and analysis than previously provided. Furthermore, we present new ground-based data obtained with the 1.0-m and 1.9-m telescopes of the South African Astronomical Observatory (SAAO), and the Las Cumbres Observatory \citep[LCO;][]{brown13} 1.0-m telescope network sited at the Siding Springs Observatory (SSO), Australia, and the Cerro Tololo Interamerican Observatory (CTIO), Chile.

\subsection{SuperWASP data}
\label{WASP}

SuperWASP is a leading ground-based project in the search for transiting exoplanets. It also produces excellent data for the study of stellar variability. For full details of the SuperWASP project, see \citet{pollacco06}; for examples of stellar variability studies that have been conducted with the data, we refer the reader to the following papers: \citet{maxted08,thomas10,norton11,norton16,smalley11,smalley14b,smalley17,holdsworth14a,holdsworth17a,bowman15,greer17}.

SuperWASP observed J1640 over three years, 2010, 2011 and 2012. Standard data processing is applied to the data on ingest to the archive. On retrieval of the data, out-lying points are discarded using a resistant mean algorithm. This process serves to minimise the noise of the amplitude spectrum (see \citealt{holdsworth14a} for an example and details). The final light curve which we present here consists of  20\,567 data points. Details of the WASP observations, and frequency analysis, are shown in Table\,\ref{tab:wasp}.

\begin{table*}
    \caption{Log of the WASP observations and frequency analysis results. BJD is the start time of the observations, given as BJD-245\,0000.0. A non-linear least-squares fit was used to determine the frequencies and amplitudes.}
\label{tab:wasp}
  \begin{tabular}{lcrccc}
    \hline
    \multicolumn{1}{c}{Year} & BJD & \multicolumn{1}{c}{Length} & Number of & Frequency & Amplitude \\
                                                &  & \multicolumn{1}{c}{(d)}        & points        &  (\cd)          & (mmag)    \\
    \hline
    2010					& 5332.3701 & 93.0234 & 8474 & $151.9343\pm0.0005$ & $3.63\pm0.32$ \\
    2011					& 5647.4961 & 146.8823 & 6803 & $151.9348\pm0.0004$ & $3.32\pm0.40$   \\
    2012					& 6012.4990 & 94.8999 & 5290 & $151.9338\pm0.0007$& $3.83\pm0.47$  \\
    \hline
  \end{tabular}
\end{table*}

The three years of SuperWASP observations provide us with a suitably long duration with which to measure the rotation period of the star. To that end, we combine all WASP data and calculate the Fourier transform to $1$\,\cd. We find the dominant peak in the amplitude spectrum at $0.27209\pm0.00004$\,\cd, equivalent to a period of $3.6753\pm0.0005$\,d. The frequency error quoted here is the formal error in the least squares-fitting, as calculated following \citet{montgomery99}. However, as the authors discuss, formal frequency errors calculated in the presence of correlated `pink' noise in the data can be unrealistically low as a) the calculation assumes white noise and b) the frequency error is proportional to the amplitude error. Therefore, by comparing the noise levels (i.e. amplitude error) in the amplitude spectrum at high frequencies -- where the noise is white -- and at the low frequency of the rotation signal, we can correct for the underestimation in the frequency error at low frequency. In doing so, we find that the real error in frequency is a factor of 1.48 higher than the formal error. Therefore, the rotation period is determined to be $3.6753\pm0.0008$\,d. We detect no harmonics of the rotation period in the WASP data. The amplitude spectrum and phase folded light curve of all WASP data are shown in Fig\,\ref{fig:lc}.

\begin{figure}
\includegraphics[width=\linewidth]{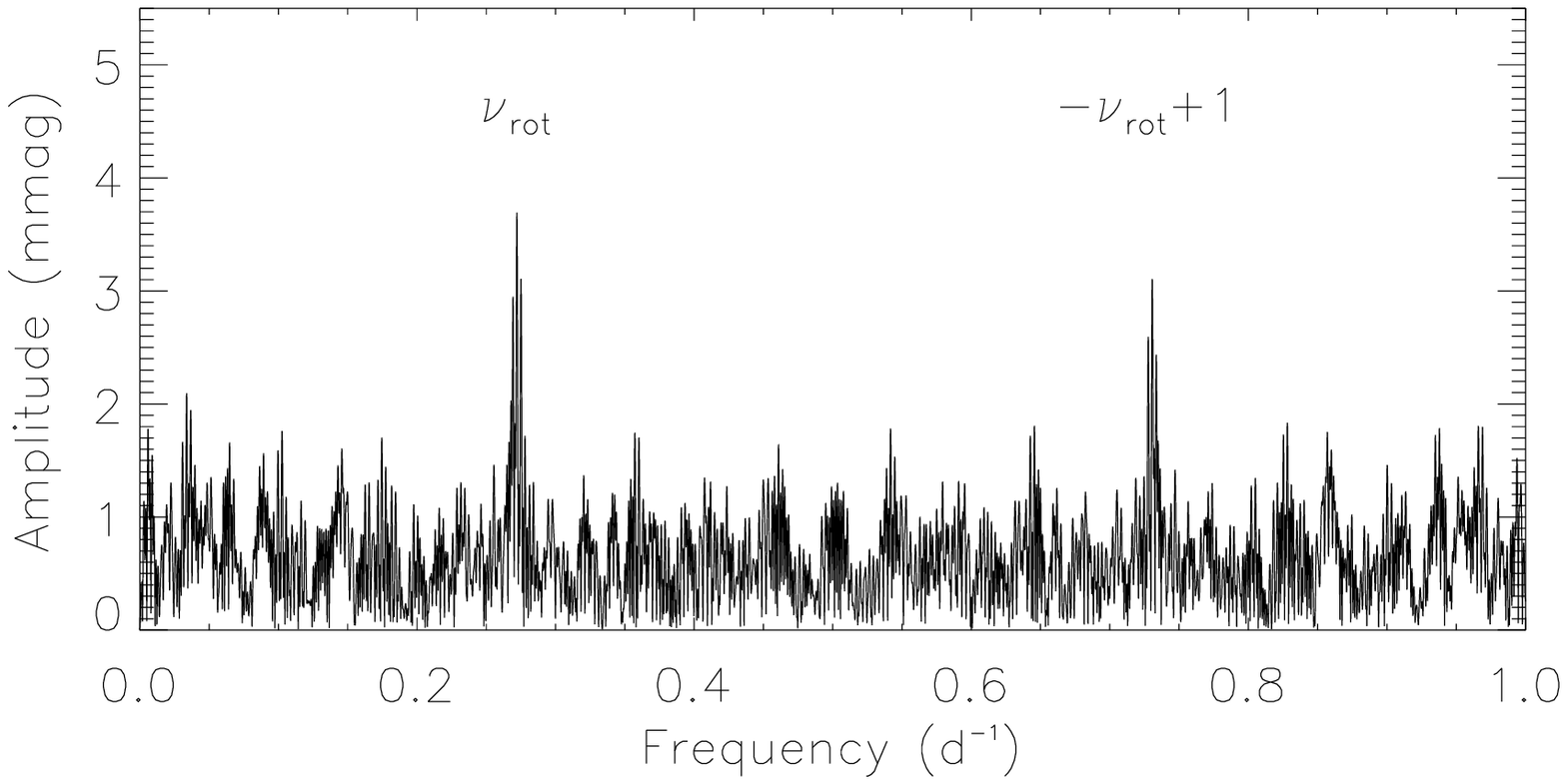}
\includegraphics[width=\linewidth]{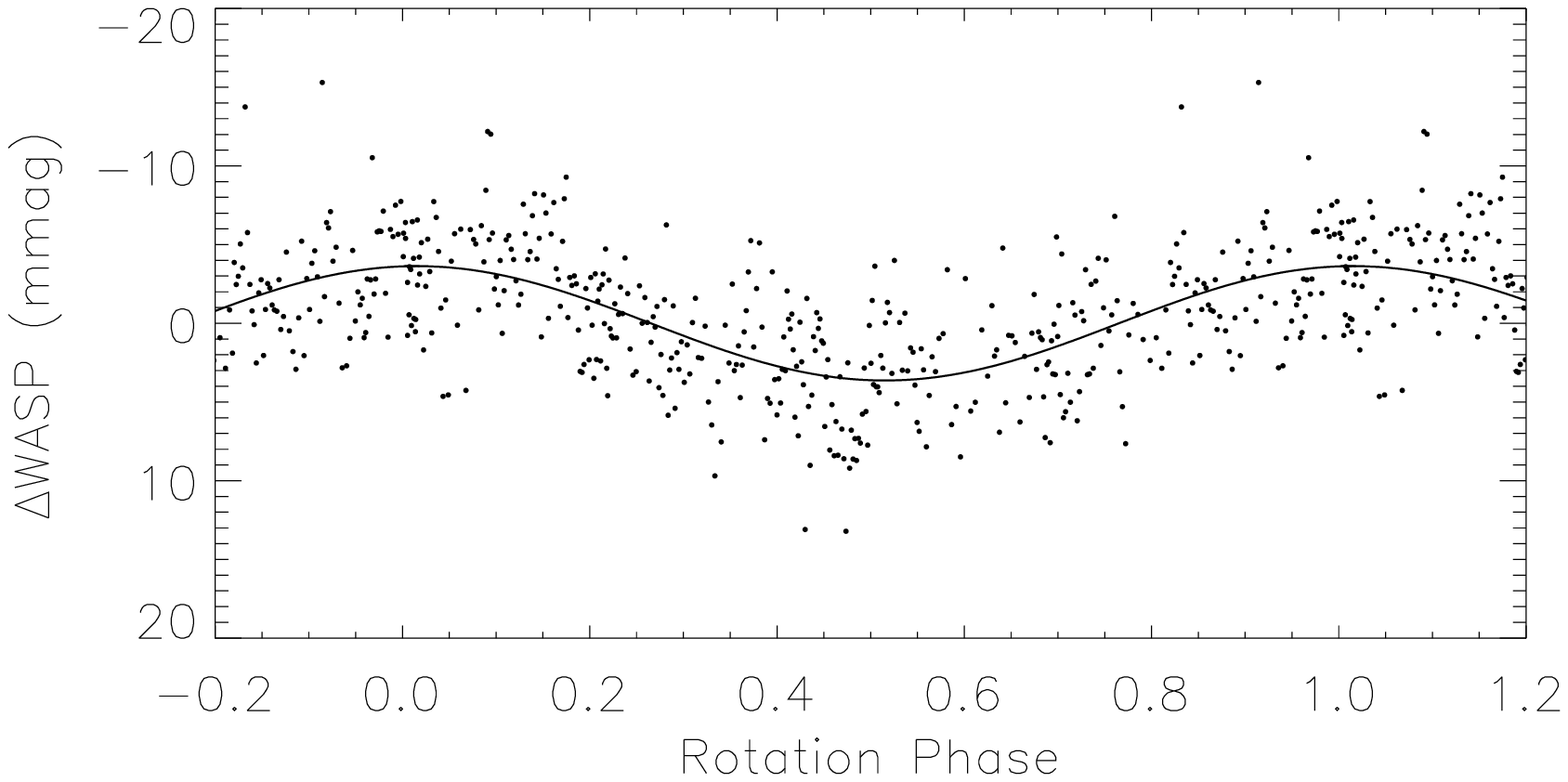}
  \caption{Top: low-frequency amplitude spectrum of the WASP light curve showing the rotation frequency of J1640, as well as the $+1$\,\cd\, alias of the negative component. Bottom: the WASP light curve folded on the rotation frequency, i.e. $\nu=0.27209$\,\cd. The data are singly periodic with a period of $1/\nu=3.6753$\,d. The data have been binned 50:1. }
  \label{fig:lc}
\end{figure}

To analyse the high-frequency variations in the WASP data, we pre-whiten each individual year in the frequency range $0 - 10$\,\cd\, to the noise level in the frequency range around the pulsation. This procedure removes the rotation signature as well as any remaining systematic noise that is not fully removed by the SuperWASP pipeline. We then calculate the frequency and amplitude of the pulsation, fit the results using non-linear least-squares with the results show in the last two columns in Table\,\ref{tab:wasp}. The frequencies and amplitudes are all in agreement, within the errors, over the three years of observations; the pulsation is apparent at a frequency of $151.934$\,\cd\, with an average (weighted) amplitude of $3.58$\,mmag. Fig.\,\ref{fig:ft} shows an amplitude spectrum of all the WASP data combined.

\begin{figure}
  \includegraphics[width=\linewidth]{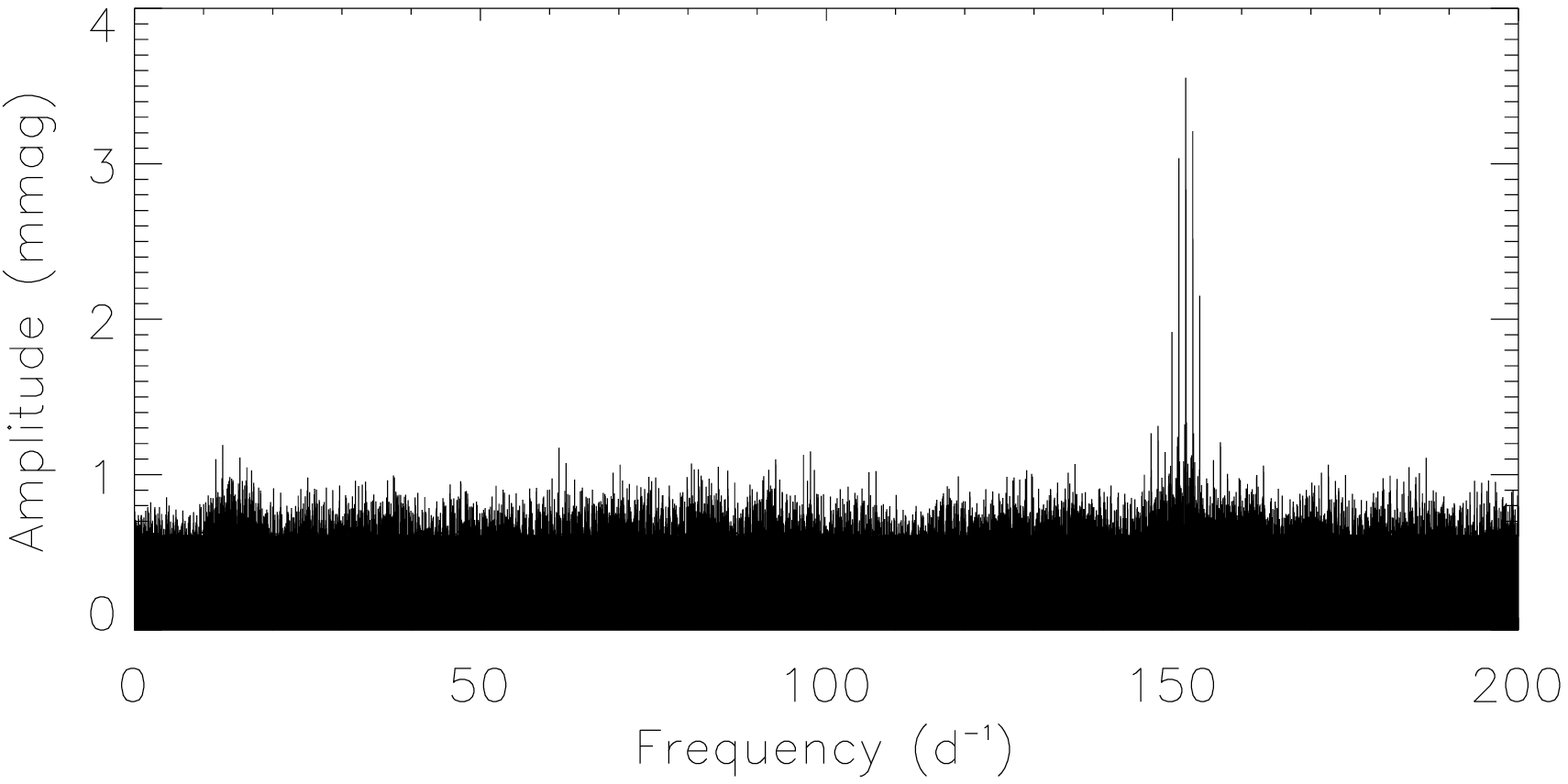}
  \caption{Amplitude spectrum of all WASP data. Low frequency signals have been removed between $0 - 10$\,\cd. The structure surrounding the pulsation signature is a result of daily aliases.}
  \label{fig:ft}
\end{figure}

We can extract no further information from the WASP data as the noise dominates over any sidelobes of the pulsation, or further, low-amplitude, pulsations. 

The amplitudes presented in this section are those observed with the SuperWASP broadband filter (i.e. $4000-7000$\,\AA). Observations of the roAp stars are typically made with $B$ filters where the pulsational signal-to-noise ratio is greatest. The amplitude reduction in the SuperWASP filter is expected to be of the order 2.3 following the comparison of WASP and $B$ data of the roAp star J1940 \citep{holdsworth18a}, demonstrating that J1640 is one of the highest amplitude roAp stars yet observed.


\subsection{Follow-up observations}

J1640 was observed as a secondary target at the SAAO during 2016 June. Observations were made with the Sutherland High Speed Optical Cameras \citep[SHOC; ][]{coppejans13} mounted on both the 1.0-m and 1.9-m telescopes. We used a Johnson $B$ filter for all observations. Integration times were either 5 or 10\,s with a readout time of 6.7\,ms. Table\,\ref{tab:log} shows a log of the observations.  

\begin{table*}
    \caption{Details of the follow-up observations of J1640, where BJD$=$BJD-245\,0000.0 and is the time stamp of the first data point.}
\label{tab:log}
  \begin{tabular}{llcrrrrr}
    \hline
     Year & \multicolumn{1}{c}{UTC Date} & BJD & \multicolumn{1}{c}{Length} & \multicolumn{1}{c}{Number of} & \multicolumn{1}{c}{Exposure} & \multicolumn{1}{c}{Site/Telescope} & \multicolumn{1}{c}{Observer(s)} \\ 
             &    			              &   & \multicolumn{1}{c}{(min)}   & \multicolumn{1}{c}{points}         &  \multicolumn{1}{c}{time (s)}    &		                                         &		        \\
    \hline
   2016	\\
   		& Jun 01/02	& 7541.3176	& 121.4	& 729	&	10	&	SAAO 1.0-m	&	DLH/RRS \\
		& Jun 02/03	& 7542.3481	& 115.1	& 664	&	10	&	SAAO 1.0-m	&	DLH/RRS \\
		& Jun 05/06	& 7545.3128	& 105.1	& 631	&	10	&	SAAO 1.0-m	&	DLH/RRS \\
		& Jun 06/07	& 7546.3306	& 83.4	& 501	&	10	&	SAAO 1.0-m	&	DLH/RRS \\
		& Jun 07/08	& 7547.3105	& 102.2	& 614	&	10	&	SAAO 1.0-m	&	DLH/RRS \\
		& Jun 08/09	& 7548.3355	& 92.7	& 1112	&	5	&	SAAO 1.9-m	&	DLH \\
		& Jun 11/12	& 7551.3885	& 36.8	& 222	&	10	&	SAAO 1.9-m	&	DLH \\
		& Jun 12/13	& 7552.3442	& 51.4	& 301	&	10	&	SAAO 1.9-m	&	DLH \\
		& Jun 16/17	& 7556.2825	& 112.2	& 674	&	10	&	SAAO 1.0-m	&	DLH \\
		& Jun 17/18	& 7557.2768	& 118.5	& 681	&	10	&	SAAO 1.0-m	&	DLH \\
		& Jun 18/19	& 7558.2819	& 105.7	& 635	&	10	&	SAAO 1.0-m	&	DLH \\
		& Jun 21/22	& 7561.2636	& 104.1	& 625	&	10	&	SAAO 1.0-m	&	DLH \\
   2017	\\
  		& May 24/25	& 7898.3613 	& 105.4	& 316	&	20	&	SAAO 1.0-m	&	DLH/DMB \\ 		 		
  		& May 27/28	& 7901.3259	& 483.1	& 1630	&	10	&	SAAO 1.0-m	&	DLH/DMB \\ 
  		& May 28/29	& 7902.3248 	& 444.6	& 2626	&	10	&	SAAO 1.0-m	&	DLH/DMB \\ 
  		& May 29/30	& 7903.3217	& 458.4	& 2735	&	10	&	SAAO 1.0-m	&	DLH/DMB \\ 
		& May 31		& 7904.5902	& 198.8	& 197	&	20	&	CTIO 1.0-m	&	Service \\
		& May 31		& 7904.9779	& 198.2	& 197	&	20	&	SSO 1.0-m	&	Service \\
  		& May 31/01	& 7905.3170	&  426.5	& 2263	&	10	&	SAAO 1.0-m	&	DLH/DMB \\ 
		& Jun 01		& 7905.5898	& 198.0	& 190	&	20	&	CTIO 1.0-m	&	Service \\	
  		& Jun 01/02	& 7906.3121	& 367.9	& 1873	&	10	&	SAAO 1.0-m	&	DLH/DMB \\ 
		& Jun 02		& 7906.5902	& 198.6	& 200	&	20	&	CTIO 1.0-m	&	Service \\	
		& Jun 02		& 7906.9777	& 110.3	& 110	&	20	&	SSO 1.0-m	&	Service \\	
  		& Jun 02/03	& 7907.3078	& 446.4	& 2877	&	10	&	SAAO 1.0-m	&	DLH/DMB \\ 
		& Jun 03		& 7907.9775	& 128.7	& 128	&	20	&	SSO 1.0-m	&	Service \\	
		& Jun 04		& 7908.5701	& 199.1	& 191	&	20	&	CTIO 1.0-m	&	Service \\	
		& Jun 04		& 7908.9775	& 38.4	& 39		&	20	&	SSO 1.0-m	&	Service \\			
  		& Jun 04/05	& 7909.3232	& 369.8	& 1175	&	10	&	SAAO 1.0-m	&	DLH/DMB \\
  		& Jun 05/06	& 7910.2994	& 426.3	& 2463	&	10	&	SAAO 1.0-m	&	DLH/DMB \\ 
		& Jun 07		& 7911.0555	& 144.4	& 143	&	20	&	SSO 1.0-m	&	Service \\	
  		& Jun 08/09	& 7913.3718 	& 223.1 	& 601	&	20	&	SAAO 1.0-m	&	DLH/DMB \\ 
		& Jun 09		& 7913.5902	& 239.2	& 229	&	20	&	CTIO 1.0-m	&	Service \\	
		& Jun 10		& 7914.6312	& 202.7	& 201	&	20	&	CTIO 1.0-m	&	Service \\	
		& Jun 11		& 7915.6571	& 200.8	& 201	&	20	&	CTIO 1.0-m	&	Service \\	
  		& Jun 11/12	& 7916.2952 	& 114.7 	& 218	&  10/20	&	SAAO 1.0-m	&	DLH/DMB \\ 
		& Jun 12		& 7916.6422	& 39.9	& 42		&	20	&	CTIO 1.0-m	&	Service \\	
  		& Jun 12/13	& 7917.2801	& 482.7	& 2878	&	10	&	SAAO 1.0-m	&	DLH/DMB \\ 
		& Jun 13		& 7917.6109	& 117.7	& 119	&	20	&	CTIO 1.0-m	&	Service \\	
  		& Jun 13/14	& 7918.3056	& 460.5	& 2180	&	10	&	SAAO 1.0-m	&	DLH/DMB \\ 
  		& Jun 21/22	& 7926.2730	& 180.3	& 1015	&	10	&	SAAO 1.0-m	&	RRS \\ 
  		& Jun 23/24	& 7928.2544	& 425.1	& 2518	&	10	&	SAAO 1.0-m	&	RRS \\ 
  		& Jun 24/25	& 7929.2826	& 384.8	& 2304	&	10	&	SAAO 1.0-m	&	RRS \\ 
  		& Jun 25/26	& 7930.2638	& 411.1	& 2460	&	10	&	SAAO 1.0-m	&	RRS \\ 
  		& Jun 29/30	& 7934.2695	& 373.2	& 2158	&	10	&	SAAO 1.0-m	&	 MJ/TL\\
  		& Jun 30/01	& 7935.2440	& 422.3	& 2339	&	10	&	SAAO 1.0-m	&	 MJ/TL\\
  		& Jul 01/02	& 7936.2289	& 395.3	& 2519	&	10	&	SAAO 1.0-m	&	 MJ/TL\\
  		& Jul 02/03	& 7937.2258	& 427.5	& 2070	&	10	&	SAAO 1.0-m	&	 MJ/TL\\
  		& Jul 03/04	& 7938.2285	& 421.2	& 2519	&	10	&	SAAO 1.0-m	&	 MJ/TL\\ 
  		& Jul 04/05	& 7939.2208	& 421.1	& 2518	&	10	&	SAAO 1.0-m	&	 MJ/TL\\ 
		& Jul 08		& 7942.0215	& 173.9	& 165	&	20	&	SSO 1.0-m	&	Service \\	
  		& Jul 19/20	& 7954.1995	& 84.2	& 502	&	10	&	SAAO 1.0-m	&	 DLH\\ 
  		& Jul 20/21	& 7955.1928	& 90.4	& 538	&	10	&	SAAO 1.0-m	&	 DLH\\ 
  		& Jul 21/22	& 7956.1851	& 120.0	& 720	&	10	&	SAAO 1.0-m	&	 DLH\\ 
  		& Jul 22/23	& 7957.1911	& 94.5	& 540	&	10	&	SAAO 1.0-m	&	 DLH\\ 
  		& Jul 23/24	& 7958.1861	& 90.2	& 538	&	10	&	SAAO 1.0-m	&	 DLH\\ 
  		& Jul 24/25	& 7959.1964	& 75.0	& 450	&	10	&	SAAO 1.0-m	&	 DLH\\ 
  		& Jul 26/27	& 7961.1917	& 59.9	& 360	&	10	&	SAAO 1.0-m	&	 DLH\\ 
  		& Jul 28/29	& 7963.1899	& 59.9	& 360	&	10	&	SAAO 1.0-m	&	 DLH\\ 
  		& Jul 29/30	& 7964.2092	& 150.6	& 899	&	10	&	SAAO 1.0-m	&	 DLH\\ 
  		& Jul 30/31	& 7965.1880	& 59.9	& 360	&	10	&	SAAO 1.0-m	&	 DLH\\ 
\\
Total 		&			&			& 13093 	& 61263	&		&				&		\\

    \hline
  \end{tabular}
\end{table*}

Subsequently, the star was observed in 2017 May, June and July, again with the 1.0-m SAAO telescope, as well as with the LCO 1.0-m telescope network located in Chile and Australia, also through a $B$ filter. These facilities are listed as CTIO and SSO in Table\,\ref{tab:log}, respectively.

The SAAO data are reduced using a customised pipeline in {\sc python}. Each data cube (typically 1\,hr long) consists of 360 image frames each with a 10\,s exposure time and 6.7\,ms readout time. Each image frame is bias corrected and normalised by a flat field. Aperture photometry is performed using optimised elliptical apertures determined from growth curves of total aperture flux versus aperture area, which capture changing values of FWHM for each source caused by variable seeing. An aperture with the same area as the target star aperture is placed nearby on the CCD, where no sources are present, to determine and subtract the sky background. This is also performed for the comparison star. Time stamps from the original {\sc fits} headers are converted to Barycentric Dynamical Time (TDB) and expressed in Barycentric Julian Date (BJD-245\,0000.0).

The LCO data are provided as fully reduced data using their in-house pipeline, {\sc{banzai}}\footnote{\url{https://lco.global/observatory/data/BANZAIpipeline/}}. Therefore, we use those frames and conduct aperture photometry, and time stamp generation, as described above. 

In the following analysis, we bin all data to the equivalent of a 20-s exposure time to ensure that all data points have the same weight. 

\subsubsection{Rotation signature}
\label{sec:SAAO_rot}
To analyse the low-frequency rotation signature, we perform differential photometry to remove the seeing, airmass, and transparency effects. The comparison star is $\sim105$\,arcsec from the target star with a magnitude of $V=13.1$. Colour indices imply this star to be a late G or early K star, hence it may also show signatures of rotational variability. However, as we know details of the rotation period of J1640 from WASP data, this possible contamination is not an issue.

As with the WASP data, we calculate a Fourier transform in the frequency range $0 - 1$\,\cd\, and find the dominant frequency to be $0.272130\pm0.000006$\,\cd, corresponding to a period of $3.67471\pm0.00008$\,d. As before, we increase our errors to provide more realistic errors in the presence of correlated noise, in this case by a factor of $5.95$, providing a rotation period of $3.6747\pm0.0005$\,d. This value is in agreement with the WASP value, an expected result due to the long term stability of spots on Ap stars.

The amplitude of modulation in the $B$ observations is much larger than in the WASP data, at $19.96$\,mmag, but even with this greater signal-to-noise, we do not detect any harmonics of the rotation period. A phase folded light curve with this result is shown in Fig.\,\ref{fig:SAAO_lc}. Note that the amplitude varies in anti-phase with the WASP data. An anti-phase variation is often seen in Ap stars, though with observations in the visual and ultraviolet. It is thought that flux redistributed from rare-earth element line-blanketing causes this effect between the two passbands \citep{molnar73}. However, it is less common to see variations between $B$ and $V$ (or the broadband WASP filter) \citep[e.g.][]{kurtz86,grobel17,drury17}, which indicates a null wavelength in the visual part of the spectrum \citep{molnar75}. 

\begin{figure}
\includegraphics[width=\linewidth]{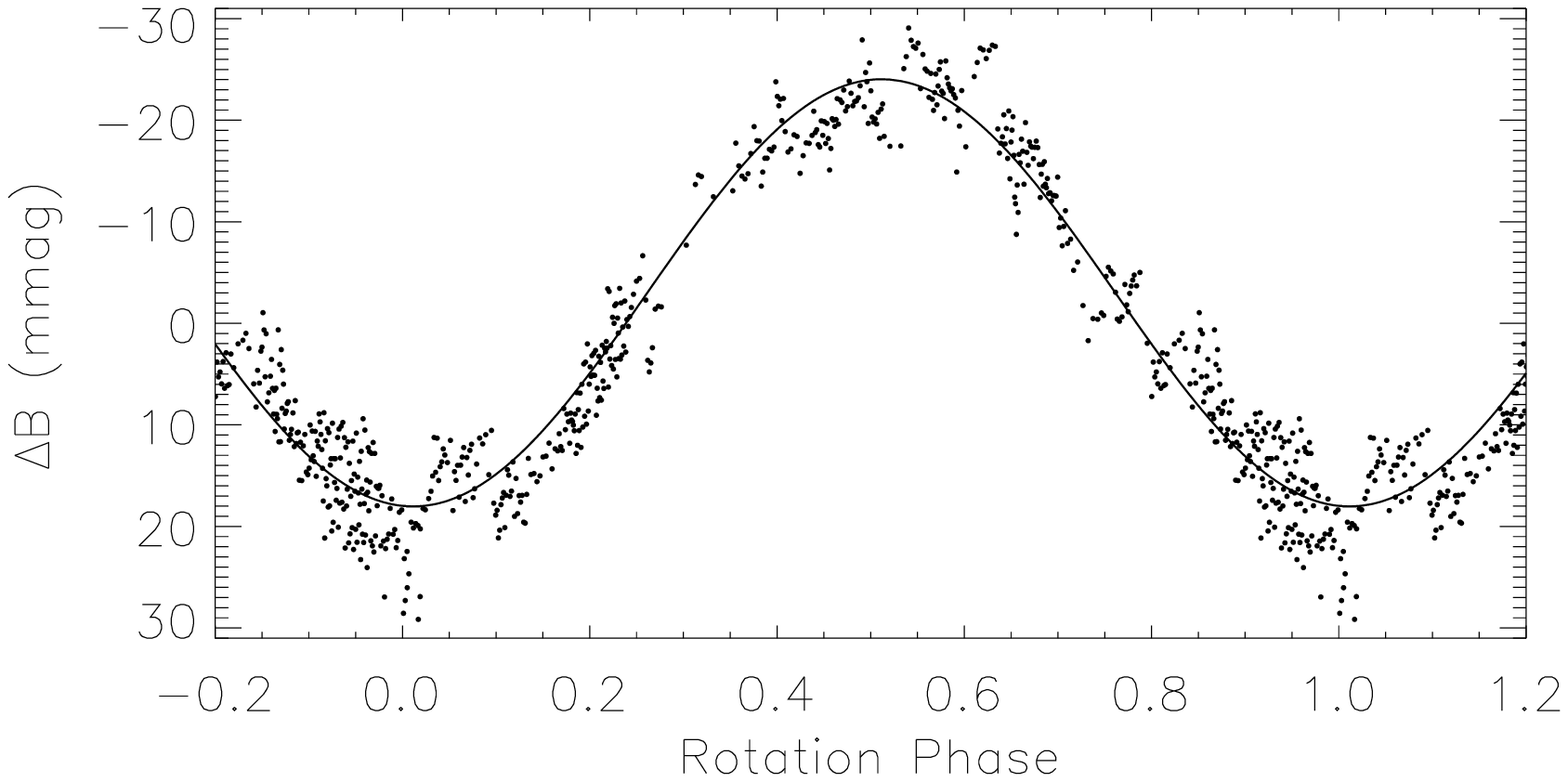}
  \caption{SAAO and LCO phased light curve folded on the rotation period, i.e. $3.6747$\,d. The plot shows a clear sinusoidal signature. The data have been binned 50:1. Note that the variation is in anti-phase with the white-light WASP observations.}
  \label{fig:SAAO_lc}
\end{figure}

\subsubsection{Pulsation signature}

We use non-differential photometry to analyse the pulsation in J1640 as this provides us with more precise data to work with, having avoided combining photon statistical observational errors of the comparison star with those of the target star when calculating the differential magnitudes. We fit and remove the airmass effects and other low-frequency variability, such as the rotation, in the range $0 - 10$\,\cd\ to the  noise level at high frequency. Using non-differential photometry, however, does mean the loss of some data on non-photometric nights. After removing obvious outlying points (and those affected by cloud), and binning the data to 20\,s integrations, the following analysis is conducted using 25\,302 data points. The full light curve used for the pulsation analysis is shown in Fig\,\ref{fig:full_lc}. 

\begin{figure*}
\includegraphics[width=\linewidth]{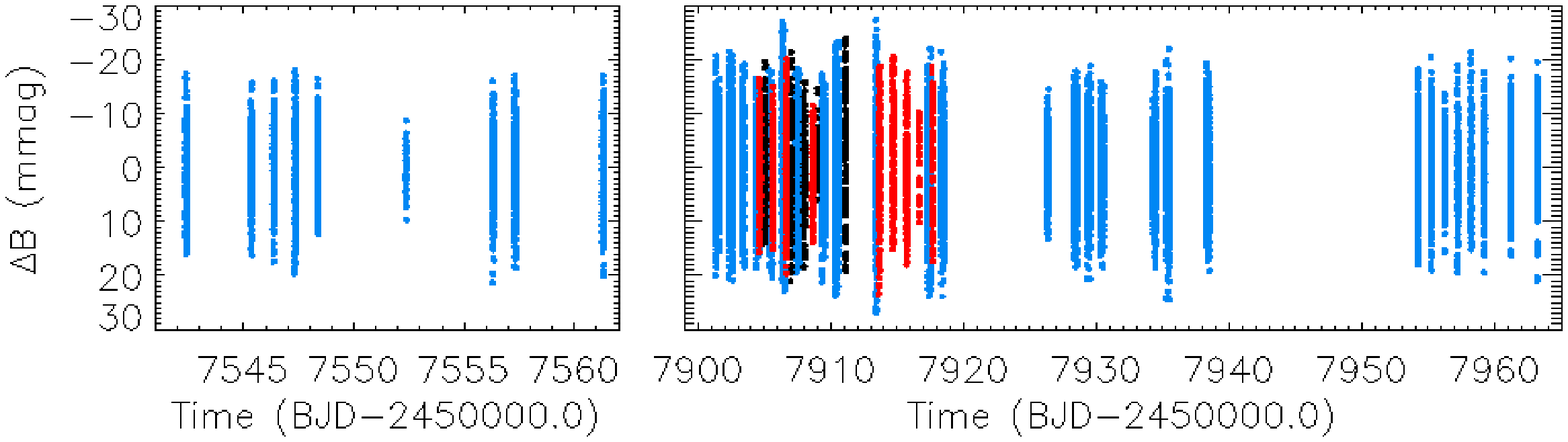}
  \caption{Full light curve of J1640 which is used for the pulsation analysis. Different colours represent different observatories: blue=SAAO; red=LCO-CTIO; black=LCO-SSO. The data are split into two separate boxes to represent the 2016 data (left) and the 2017 data (right).}
  \label{fig:full_lc}
\end{figure*}

We calculate an amplitude spectrum of the full data set which is shown in Fig.\,\ref{fig:full-ft}; the pulsation frequency is clearly evident. Fig.\,\ref{fig:ft-pw} shows a zoom of the pulsation frequency (top panel), and subsequent pre-whitening steps of the pulsation and sidelobes. After identifying and extracting the pulsation frequency, we extract the sidelobes in pairs by identifying the dominant peaks around their expected position. This was an obvious procedure for the $\pm1\nu_{\rm rot}$ pair (as shown in the second panel of Fig.\,\ref{fig:ft-pw}), but a more challenging task for the $\pm2\nu_{\rm rot}$ pair (shown in the third panel of Fig.\,\ref{fig:ft-pw}). The presence of the sidelobes, split by the previously derived rotation frequency, within the errors, enables us to confirm that we have calculated the correct rotation frequency of the star, i.e. we have identified the true rotation period in Section\,\ref{sec:SAAO_rot} and not a harmonic of it.

We fit the pulsation signature in the light curve with non-linear least-squares, and show the results in Table\,\ref{tab:nlls}. The presence of the sidelobes in the Fourier transform imply that J1640 is a quadrupole pulsator like J1940 \citep{holdsworth18a}, or a distorted dipole pulsator similar to HD\,6532 \citep{kurtz96}. Given that the shape of the quintuplet is very different from that of the distorted dipole seen in HD\,6532, and similar to that of the distorted quadrupole of J1940, we deduce that J1640 is pulsating in a quadrupole mode. This deduction is supported by our models of the pulsation, as will be seen later in Section\,\ref{sec:modelling}.
\begin{figure}
\includegraphics[width=\linewidth]{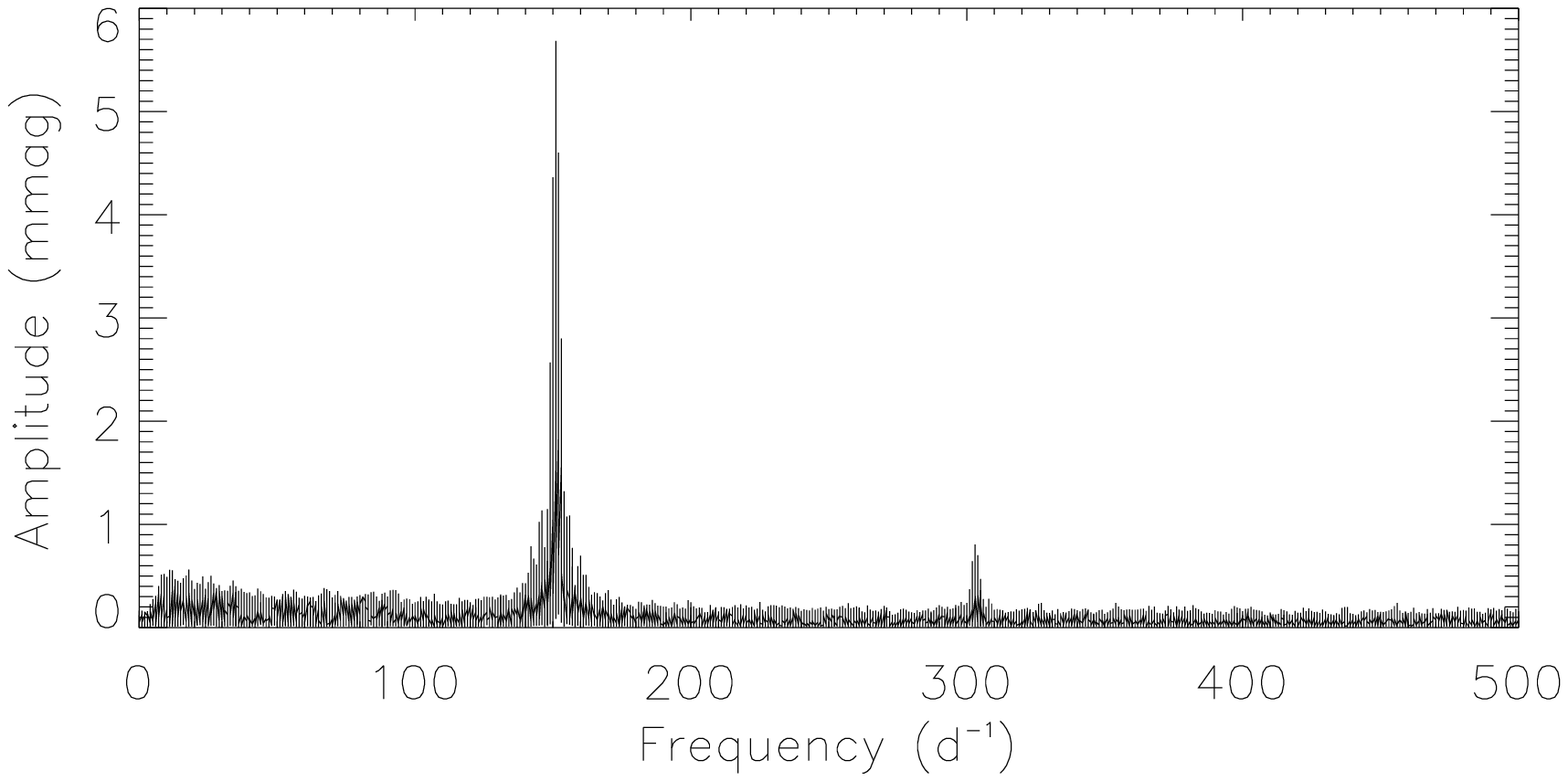}
\includegraphics[width=\linewidth]{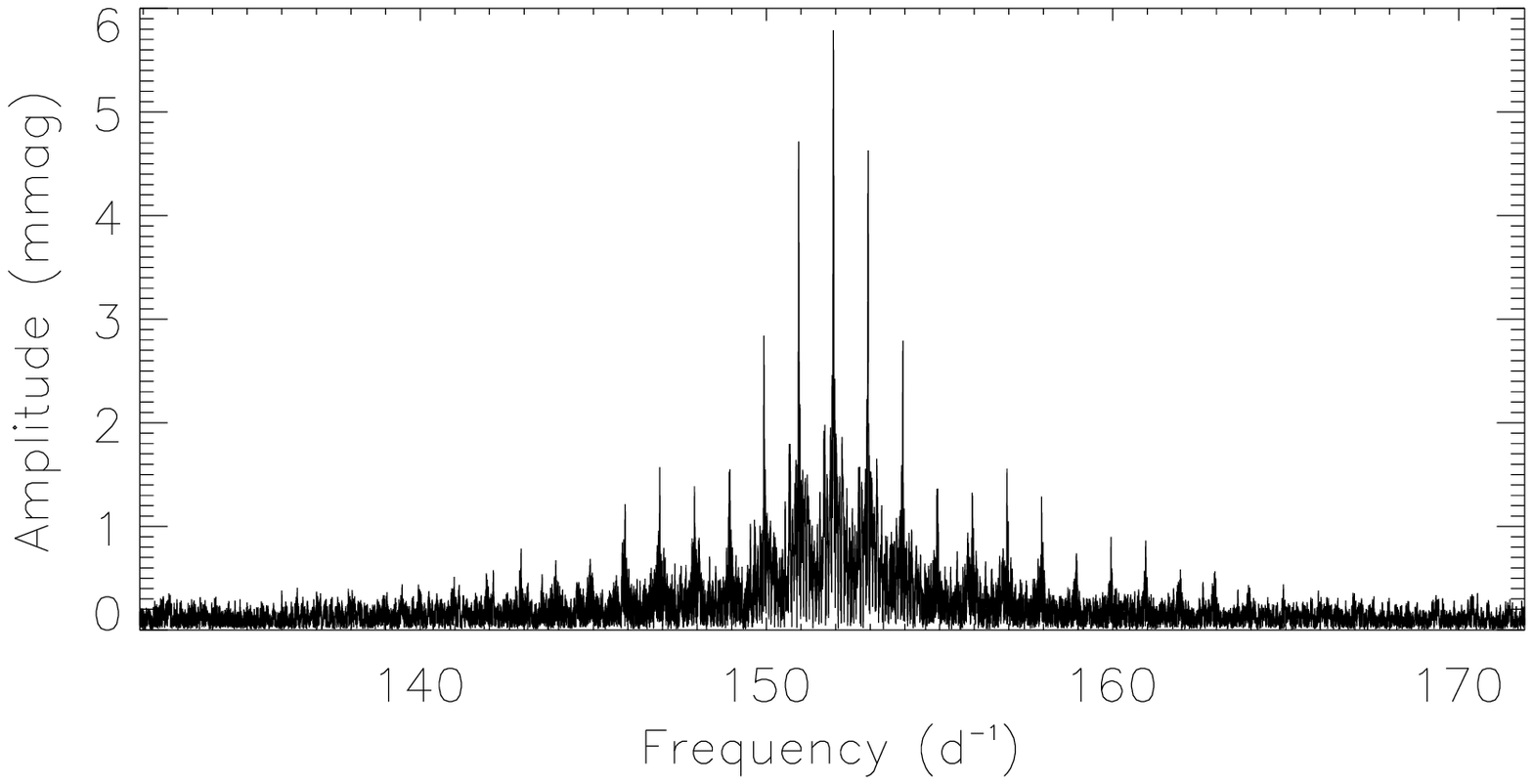}
  \caption{Top: an amplitude spectrum of the full data set showing the pulsation and the second harmonic. The third harmonic of the pulsation is present (see Table\,\ref{tab:harmonics}), but not visible on this scale. Bottom: a zoomed view of the pulsation. The surrounding structure is a result of gaps in the data, i.e. the window function.}
  \label{fig:full-ft}
\end{figure}

\begin{figure*}
\includegraphics[width=\textwidth]{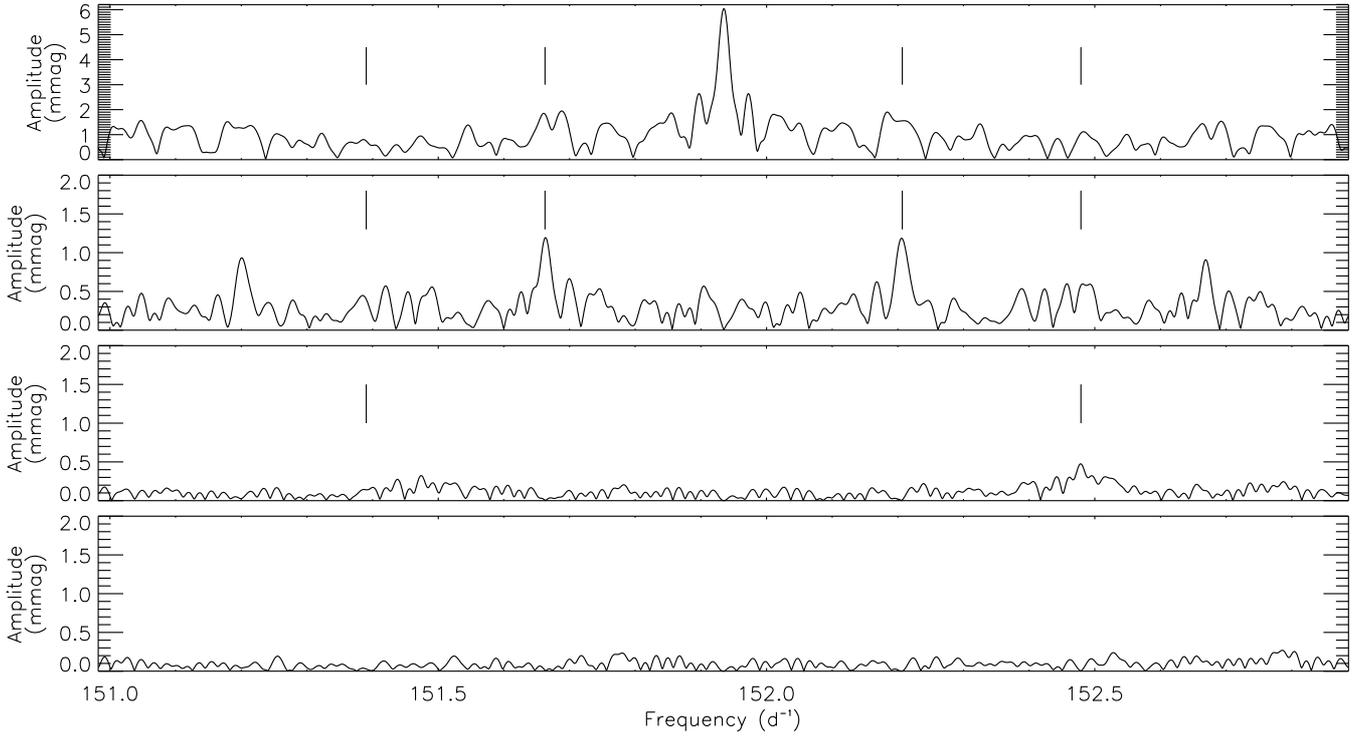}
 \caption{Zoomed plot of the pulsation and rotational sidelobes, identifying the consecutive pre-whitening of the peaks, and the amplitude spectrum of the residuals (bottom). The vertical bars indicate the expected positions of the $\pm\nu_{\rm rot}$ and $\pm2\nu_{\rm rot}$ sidelobes. The $+2\nu_{\rm rot}$ peak is clearly present, however the $-2\nu_{\rm rot}$ peak is in the noise. Note the change in amplitude scale between the top and subsequent panels.}
  \label{fig:ft-pw}
\end{figure*}

\begin{table*}
  \caption{Non-linear least-squares fit results to the full light curve. The frequency difference between one line and the previous is shown in the last column. The zero-point is BJD-245\,7753.33141. Although at low amplitude, the $\nu-2\nu_{\rm rot}$ sidelobe has a $2.6\,\sigma$ significance, thus we believe it to be real.}
  \label{tab:nlls}
  \begin{tabular}{lccrc}
    \hline
    ID & Frequency & \multicolumn{1}{c}{Amplitude} & \multicolumn{1}{c}{Phase} & Frequency difference\\
       & (\cd)     & \multicolumn{1}{c}{(mmag)}    & \multicolumn{1}{c}{(rad)} & (\cd) \\
    \hline
$\nu-2\nu_{\rm rot}$ 	&$	151.39819	\pm  0.00059	$&$  0.17	\pm  0.06	$&$  -0.500\pm0.631$&\\
$\nu-1\nu_{\rm rot}$ 	&$	151.66245	\pm	0.00008	$&$	1.17	\pm	0.06	$&$	-2.842\pm0.087	$&$0.26428\pm0.00059$\\
$\nu$              		&$	151.93463	\pm	0.00002	$&$	5.80\pm	0.06	$&$	-0.642\pm0.017	$&$0.27218\pm0.00008$ \\
$\nu+1\nu_{\rm rot}$	&$	152.20467	\pm	0.00008	$&$	1.07	\pm	0.06	$&$	2.896\pm0.090	$&$0.27004\pm0.00009$ \\
$\nu+2\nu_{\rm rot}$	&$	152.47833	\pm	0.00019	$&$	0.49	\pm	0.06	$&$	3.131\pm0.201	$&$0.27366\pm0.00021$ \\
\hline
  \end{tabular}
\end{table*}

After fitting and removing the quintuplet, we search for further, previously undetected, pulsation modes. However, we find none at the level of $0.4$\,mmag, implying J1640 pulsates in a single mode. Although we do not detect further pulsations, the harmonics of the pulsation, up to and including $3\nu$, are present in the data (although only $2\nu$ is visible at the scale of Fig.\,\ref{fig:full-ft}). The extracted harmonics are shown in Table\,\ref{tab:harmonics}. This is a common trait of the roAp stars and shows that the pulsations in these stars are non-sinusoidal; i.e., the pulsations are nonlinear. The physical cause of this nonlinearity in such low-amplitude pulsators is still unknown. 

\begin{table}
  \caption{Non-linear least-squares fit results to the full light curve, fitting just the harmonic series. The zero-point is BJD-245\,7753.33141.}
  \label{tab:harmonics}
  \begin{tabular}{lccr}
    \hline
    ID & Frequency & \multicolumn{1}{c}{Amplitude} & \multicolumn{1}{c}{Phase}\\
       & (\cd)     & \multicolumn{1}{c}{(mmag)}    & \multicolumn{1}{c}{(rad)}\\
        \hline
         $\nu$ & $151.93462\pm0.00001$ & $5.98\pm0.06$ & $-0.642\pm0.015$ \\
         $2\nu$ & $303.86916\pm0.00008$ & $1.03\pm0.06$ & $-1.128\pm0.088$ \\
 	 $3\nu$ & $455.80343\pm0.00029$ & $0.29\pm0.06$ & $-1.370\pm0.311$ \\
	\hline
  \end{tabular}
\end{table}

As previously stated, the pulsations in roAp stars can be represented using the oblique pulsator model. The model predicts the existence of sidelobes to the main pulsation frequency which are split by an exact frequency -- the rotation frequency. We test this model for J1640 by forcing the frequency of the sidelobes to be exactly split by the rotation frequency, and then fit the data with linear least-squares. When performing the fit, we select an appropriate zero-point to make the phases of the first sidelobes to be equal. Table\,\ref{tab:all-lls} shows the result of the test. If the pulsation were a pure quadrupole mode, described by a single spherical harmonic, the quintuplet phases should be equal.

\begin{table}
\centering
  \caption{Linear least-squares fit results to the full light curve. The phases of the first sidelobes have been forced to be equal by setting the zero-point to be BJD-245\,7752.30018.}
  \label{tab:all-lls}
  \begin{tabular}{lccrc}
    \hline
    ID & Frequency & Amplitude & \multicolumn{1}{c}{Phase}\\
       & (\cd)     & (mmag)    & \multicolumn{1}{c}{(rad)} \\
    \hline
$\nu-2\nu_{\rm rot}$ 	&$	151.39037 $ & $	0.05\pm	0.06	$&$1.353\pm	1.239$\\
$\nu-1\nu_{\rm rot}$ 	&$	151.66250 $ & $	1.17\pm	0.06	$&$0.884\pm	0.051$\\
$\nu$              		&$	151.93463 $ & $	5.77\pm	0.06	$&$1.376\pm	0.010$\\
$\nu+1\nu_{\rm rot}$	&$	152.20676 $ & $	1.01\pm	0.06	$&$0.884\pm	0.058$\\
$\nu+2\nu_{\rm rot}$	&$	152.47889 $ & $	0.48\pm	0.06	$&$1.082\pm	0.126$\\
\hline
  \end{tabular}
\end{table}

The phases derived from the least-squares fit are almost equal, however the average separation of the central peak and the first sidelobes are greater than $8\,\sigma$. This result confirms our conclusion that J1640 is pulsating in a distorted mode, and will be addressed in detail in Section\,\ref{sec:geom}.


\subsection{Testing amplitude and phase variability}

The stability of pulsations in roAp stars has been a topic for discussion since long-term, high-precision observations have been available; some of the stars show very stable pulsations while others show significant variability. HR\,3831 is a prime example of how a time base of 16\,yr allowed \citet{kurtz94,kurtz97} to discuss significant frequency variability. More recently, high photometric precision has allowed the for the identification of phase/frequency changes in roAp stars observed by the {\it Kepler} space telescope \citep[e.g.][]{holdsworth14b,smalley15}. 

Furthermore, the recent analysis of quadrupole roAp pulsators has found them \emph{all} to show unexpected pulsational phase variations \citep{balona11b,holdsworth14b,holdsworth16,holdsworth18a}. For a pure quadrupole mode, the oblique pulsator model predicts a phase reversal as a pulsation node crosses the line-of-sight. However, this has not been the case for these pulsators, as shown in figure\,14 of \citet{holdsworth18a}. In these cases, it has been concluded that the stars, which have been modelled in detail, are pulsating with distorted modes (for a theoretical discussion, see e.g. \citealt{shibahashi93}).

With this in mind, we test whether J1640 also shows unexpected frequency/phase variations over its rotation period, when compared to the theory for a pure normal mode. The amplitude and phase are calculated, at fixed frequency, in short sections of the light curve. These sections consist of 20 pulsation cycles, i.e. are about 0.13\,d in length, dictated by a trade off between rotation phase coverage and frequency resolution. Fig.\,\ref{fig:all-ph-amp-phi} shows the results of this test. 

 \begin{figure}
  \includegraphics[width=\linewidth]{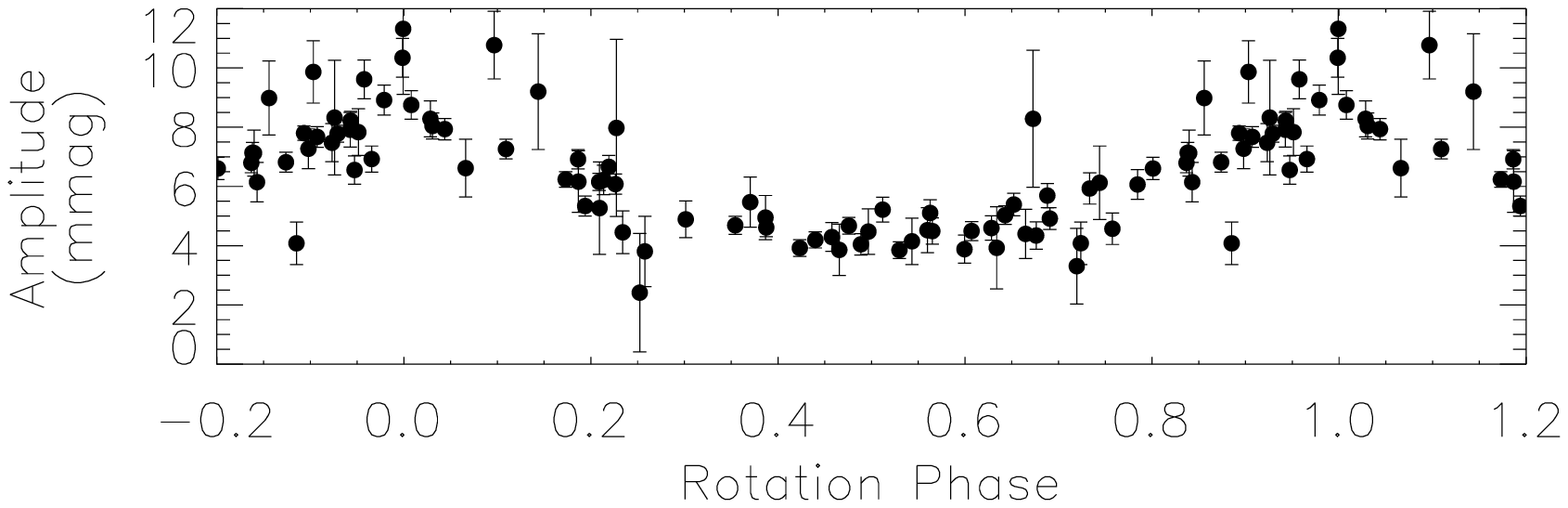}
  \includegraphics[width=\linewidth]{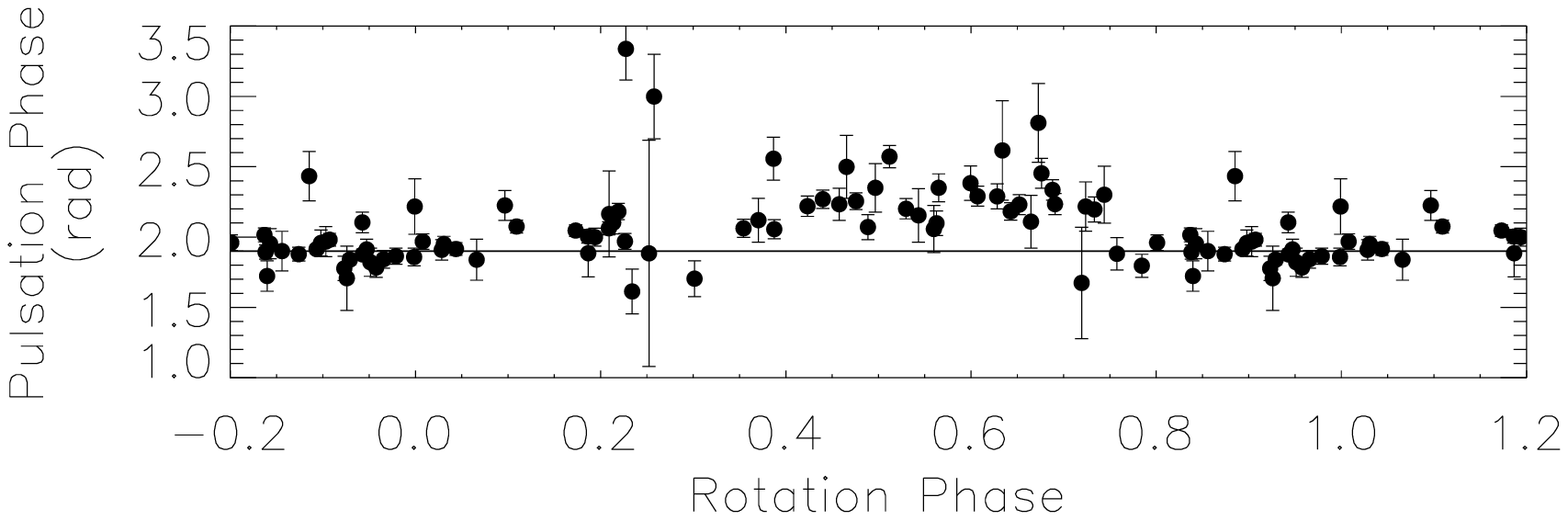}
  \caption{Top: pulsation amplitude variation with rotation phase. Bottom: variation of the pulsation phase over the rotation phase. The solid line is to guide the reader in noticing the slight phase change between about 0.3 and 0.7 in rotation phase.}
  \label{fig:all-ph-amp-phi}
\end{figure}

Considering a pure quadrupole pulsator, where the line-of-sight crosses one node, we expect the amplitude to be modulated to zero twice per rotation, and the phase to change by $\upi$\,rad. In the case of J1640, Fig.\,\ref{fig:all-ph-amp-phi} shows this does not happen. The amplitude variations seen in J1640 are reminiscent of those seen in the roAp star $\alpha$\,Cir, which is thought to pulsate in a dipole mode. In that case, as the star rotates, only one pulsation pole is seen resulting in a smooth sinusoidal variation. The smooth variation in the pulsation amplitude is a result of the change in viewing aspect of the one pulsation pole \citep{kurtz94b,bruntt09}. This casts doubt over our earlier deduction that J1640 is pulsating with a quadrupole mode.

However, in the case of $\alpha$\,Cir, the phase is also constant, as one would expect when only observing one pulsation pole. This is not what is seen here with J1640. Between approximate rotation phases of 0.3 and 0.7, there is a small bump in the pulsation phase, as shown in the bottom panel of Fig.\,\ref{fig:all-ph-amp-phi} (the solid line in the plot is to guide the eye to the small phase variation only), but this is still not the expected $\upi$\,rad change.

It is known that the rotational variation in the star's brightness generates apparent pulsation amplitude variations \citep{kurtz82}. Because of this, the pulsation amplitude will be modulated in proportion to the overall light variations. In this case, our pulsation, at 5.8\,mmag, combined with the rotation variation at 19.96\,mmag, will produce two rotational sidelobes of the order 0.06\,mmag. This is the same order as our error bars on the amplitude measurements, and much below the detected sidelobe amplitudes, implying that the rotational light variations cannot explain the presence of the observed sidelobes.

As well as the pulsation amplitude variations we must also consider the effect of surface brightness inhomogeneities on the pulsation phase, as we also test this aspect of the pulsation. As stated, for a pure quadrupole mode, the observational pulsation phase jumps by $\upi$\,rad if the difference between the integrated amplitude of the positive and negative components of the pulsation changes sign, i.e. at a certain rotational phase when the line-of-sight crosses a node; otherwise the observed pulsation phase stays constant. However, for a distorted quadrupole mode, the pulsation phase changes gradually as the ratio of the contribution of the $\ell=2$ components to the $\ell=0$ component of the amplitude changes. The presence of spots would therefore modify the ratio and hence the observational pulsation phase proportionally.  For J1640, as the spots affect the pulsation amplitude by $\sim\!1$\,pre cent, we expect comparable effects in the observed pulsation phase, i.e., much smaller than the change we observe.

This almost-constant pulsation phase is similar to some other roAp stars: KIC\,10483436 \citep{balona11b}, KIC\,7582608 \citep{holdsworth14b}, HD\,24355 \citep{holdsworth16} and J1940 \citep{holdsworth18a}. It was found that all of these stars are quadrupole pulsators pulsating with distorted modes. Given the observed phase variation, and its similarity to the other quadrupole roAp pulsators, we model J1640 in Section\,\ref{sec:modelling} to further understand the distortion of the mode.


\subsection{Mode geometry constraints}
\label{sec:geom}

Due to the misalignment of the pulsation axis and rotation axis in the roAp stars, we are able to constrain the geometry of the pulsation modes in these stars. The method of \citet{kurtz90} allows us to estimate the angle of obliquity between the rotation and the pulsation axes, $\beta$, and the inclination angle of the star, $i$ through comparing the amplitude of the rotational sidelobes. In the simplified case where the frequency quintuplet is caused by an axisymmetric, non-distorted (which it is not),  quadrupole pulsation, and neglecting limb darkening effects, the following relation can be applied:

\begin{equation}
\tan i\tan\beta=4\frac{A^{(2)}_{-2}+A^{(2)}_{+2}}{A^{(2)}_{-1}+A^{(2)}_{+1}},
\label{eq:OPM}
\end{equation}
where $i$ and $\beta$ are as above, $A^{(2)}_{\pm1}$ are the amplitudes of the first sidelobes and $A^{(2)}_{\pm2}$ are the amplitudes of the second sidelobes.

Using equation\,(\ref{eq:OPM}) and values from Table\,\ref{tab:all-lls}, we calculate $\tan i\tan\beta=0.97\pm0.16$ for this simple model; given our stated assumptions this value provides a first indication of the geometry of the mode. Due to the low resolution of the spectra presented in Section\,\ref{sec:spec}, we cannot determine $v \sin i$ and thus we are unable to disentangle $i$ and $\beta$ from the result of equation\,(\ref{eq:OPM}). However, when we model the pulsation mode in Section\,\ref{sec:modelling}, we find that values of $i=70^\circ$ and $\beta=13^\circ$ best fit the observations, resulting in $\tan i\tan\beta=0.634$. Comparing the value of $\tan i\tan\beta$ from equation\,(\ref{eq:OPM}), and the value derived using the model parameters, we can see that the simplified, non-distorted assumption is not correct.

Under the assumption that the spots causing the rotational light variations are at the magnetic poles, and that both poles have spots (which may not be the case, although is common in Ap stars), then from the single-wave rotational light variations we conclude that $i + \beta < 90^\circ$, since we see only one magnetic pole. If spots at both poles were seen, we would see a double wave rotational light variation. 

Further to this, in the case of a non-distorted quadrupole mode, a node would not pass over the line-of-sight if either $i$ and $\beta$ are near to 90$^\circ$, and we would see no phase variation of the pulsation. Therefore, for a non-distorted mode, $i-\beta$ must be less than $54.7^\circ$ for phase variations to be observed -- the co-latitude of nodes for a pure quadrupole pulsator (as given by the Legendre polynomial: $P^0_2(\cos\theta)=\frac{1}{2}[3\cos^2\theta -1]$). 

However, in the case of the model parameters, i.e. $(i,\beta)=(70^\circ,13^\circ)$, $i-\beta=57.0^\circ$. This geometry would not cause a node to cross the line-of-sight in a non-distorted mode, and would then result in no pulsational phase variation. This is similar to what we show in Fig.\,\ref{fig:all-ph-amp-phi}, however with a small change detected in the observations, indicating a distorted mode, as we have previously shown.


\section{Modelling the pulsational amplitude and phase variations}
\label{sec:modelling}

As previously stated, the oblique pulsations in roAp stars cause amplitude and phase modulations of the pulsation mode, which manifests itself as rotational sidelobes in the amplitude spectrum. When the geometry allows, the pulsation phase changes by about $\upi$~rad when a node crosses the line-of-sight, i.e. the phase changes at amplitude minima . Nevertheless, this is not the case for J1640.

As previously discussed, the presence of a frequency quintuplet in the amplitude spectrum of J1640 is not conclusive evidence that the star pulsates in a quadrupole mode, given, for example the four sidelobes of the distorted dipole pulsator HD\,6532. Furthermore, the smooth amplitude variation in J1640 is similar to that seen in $\alpha$\,Cir, also a dipole pulsator. However, given the disparity between the sidelobe phases, and the small phase variation over the rotation period of the star, we do deduce it to be pulsating in a distorted mode. 

We attempt to model this distorted pulsation by following the method of \citet{saio05}, as previously applied to model both J1940 \citep{holdsworth18a} and HD\,24355 \citep{holdsworth16}. In short, we numerically solve the eigenvalue problem for non-adiabatic linear pulsations under a dipole magnetic field, with consideration of limb darkening for the Eddington gray atmosphere, i.e. $\muup=0.6$, which is common practice for roAp stars \citep[e.g. ][]{ryabchikova97}. We expand the eigenfunction as a sum of terms proportional to $Y_\ell^0$ with $\ell = 0, 2, 4, \ldots , 38$. We refer the reader to those three papers for more detail. Here, we test both the distorted dipole and distorted quadrupole scenarios, due to our uncertainty of the mode type.

Fig.\,\ref{fig:modulation} shows the best dipole (left) and quadrupole (right) cases for where the model best replicates the amplitude and phase modulations observed in J1640. In both cases, the amplitude modulations can be reproduced well. However, in the case of the dipole model, there are no instances we can find where, when the amplitude modulation is well reproduced, the phase variations can be modelled. 

\begin{figure*}
   \includegraphics[width=0.99\columnwidth]{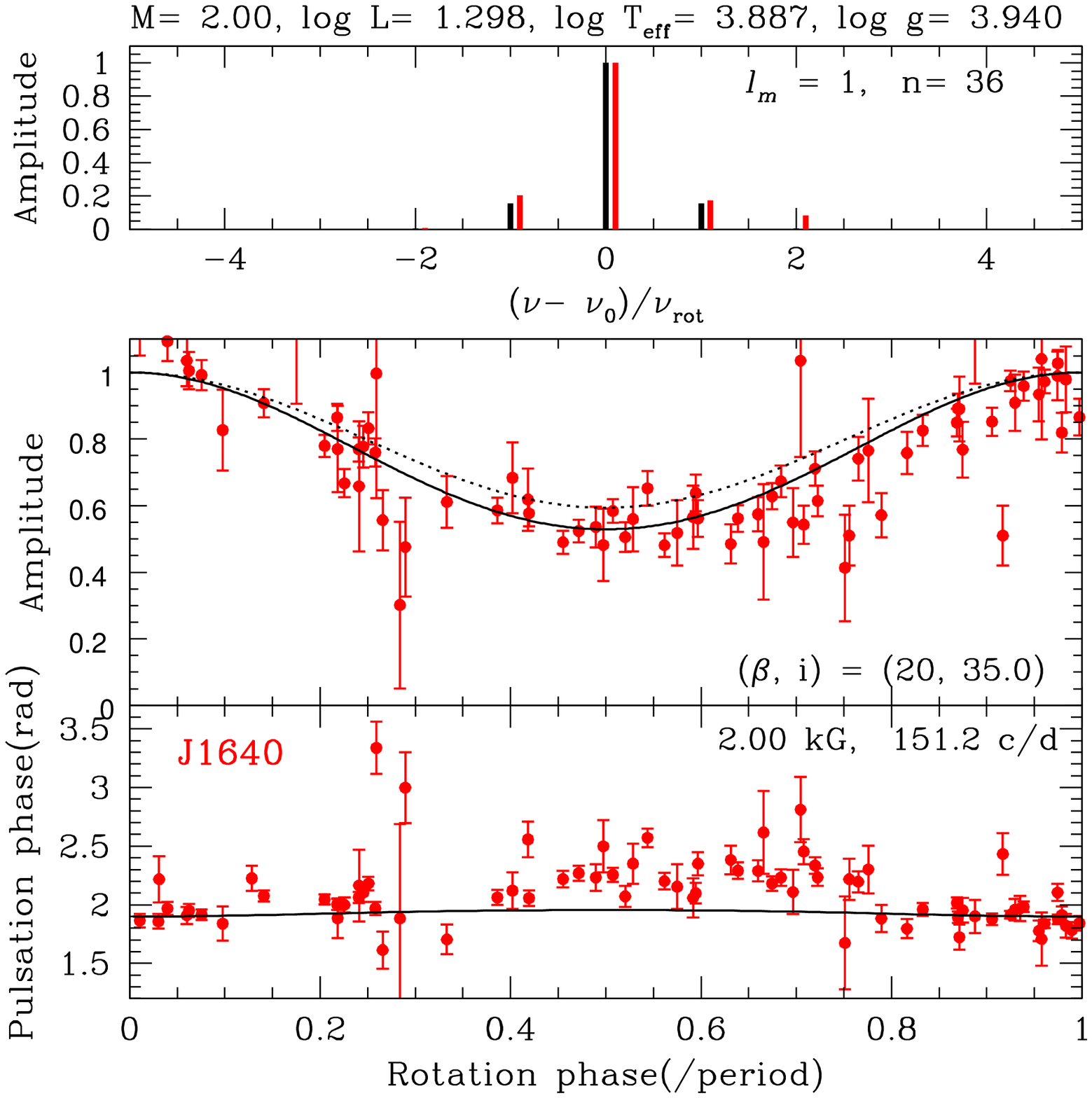}   \hfill
   \includegraphics[width=0.99\columnwidth]{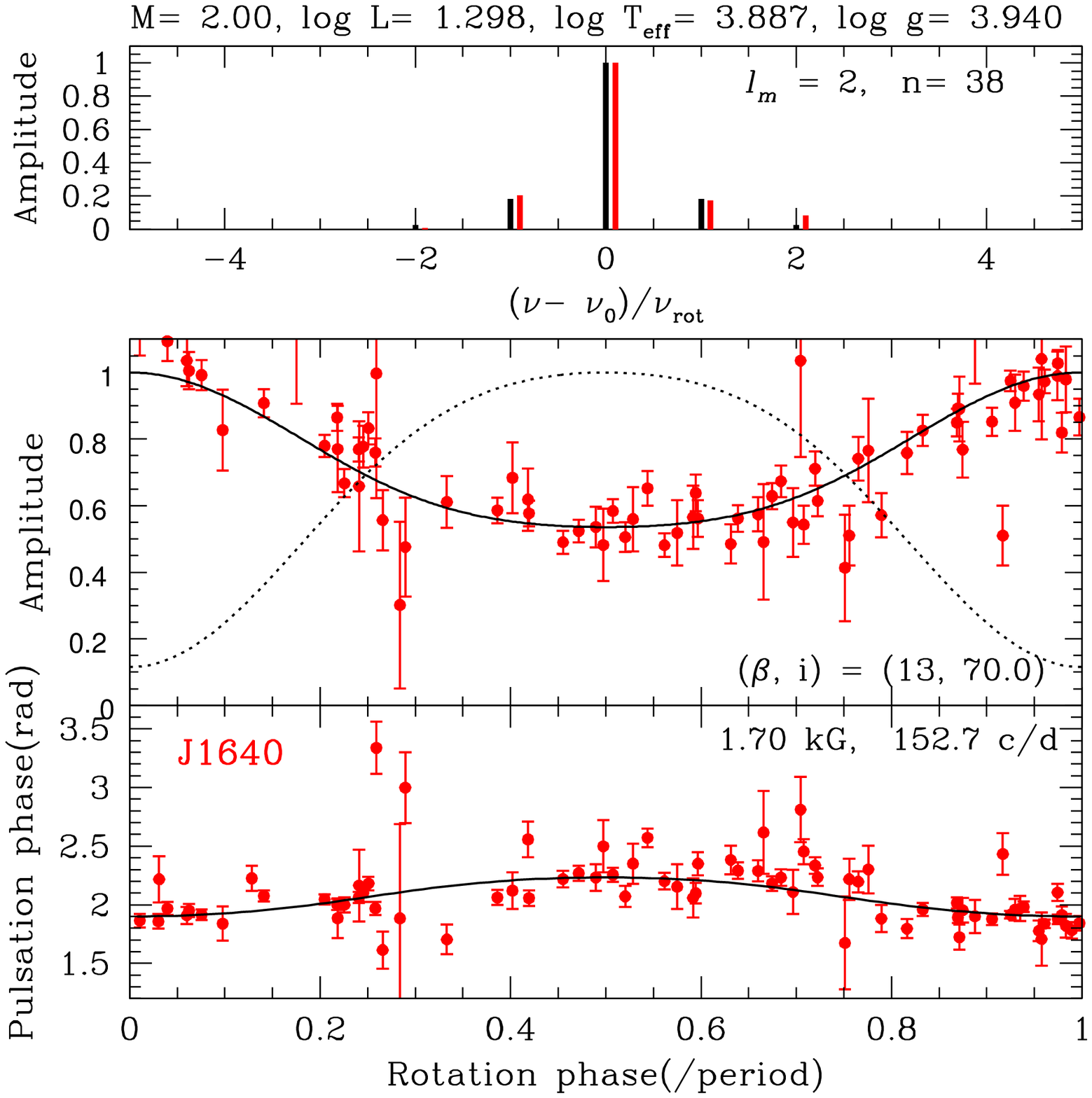}
\caption{Comparison of observed amplitude and phase modulations for the two models of J1640. Red lines/dots represent the observations, with black representing the models. Left: comparison of a distorted dipole model with the data. Right: comparison of a distorted quadrupole model with the data. In each column, the top panel shows a comparison of the quintuplet, with the observations shifted towards the right for clarity. The middle panel shows the amplitude modulations as a function of rotation phase. In each case, the best fitting theoretical distorted model is represented by the solid line, the dotted line is the expected amplitude modulation from a pure mode i.e. neglecting magnetic effects. The bottom panel shows the pulsation phase variations over the rotation cycle.}
\label{fig:modulation}
\end{figure*}

This is not the case for the distorted quadrupole model. We are able to reproduce the phase modulations simultaneously with the amplitude modulations.  Therefore, we conclude that J1640 is {\emph{most probably}} pulsating in a distorted quadrupole mode. In this case, we find the following stellar parameters are required: $M=2.0\,$M$_\odot$, $\log L/$L$_\odot = 1.30$ and a polar magnetic field strength of $B_{\rm p}=1.70$\,kG.

Here we find that the amplitude and phase modulations are produced mainly due to a combination between the spherical symmetric component ($\ell=0$) and quadrupole component ($\ell=2$), where the energy contributed to by the $\ell=0$ and $\ell=2$ components is almost equal. The small increase in pulsation phase during rotation phase between about 0.3 and 0.7 is mainly caused by the contribution of the $\ell=2$ component, which has maximum visibility between these rotation phases and an opposite pulsation phase from the $\ell=0$ component, with a small contribution from the $\ell>2$ components.

\section{Discussion}
\label{sec:disc}

To put our results into the context of other roAp stars which are pulsating in a distorted quadrupole mode, we make comparisons between KIC\,7582608 \citep{holdsworth14b}, HD\,24355 \citep{holdsworth16} and J1940 \citep{holdsworth18a}. We present a visual comparison of the four stars in Fig.\,\ref{fig:quad_comp}. We have ordered the stars, top to bottom, by the increasing contribution of the $\ell>2$ components of the spherical harmonic fit to the data. The solid lines on the figure represent the best fitting model for each star, with the parameters shown in Table\,\ref{tab:model_all}, alongside the measured frequencies and amplitudes of their respective pulsations. The amplitudes for J1640 and J1940 are the measured $B$ amplitudes. The amplitude for HD\,24355 is converted from the {\emph{Kepler} passband, Kp, to $B$ band using the relation derived by \citet{holdsworth16} where simultaneous observations of the same star were made. The amplitude for KIC\,7582608 is converted from the measured super-Nyquist amplitude in {\emph{Kepler}} Long Cadence data to an undiluted amplitude using equation\,(1) of \citet{murphy12} (due to severe under sampling of the pulsation), and then to a $B$ amplitude using the same method as for HD\,24355. The amplitudes presented in Table\,\ref{tab:model_all} are those used to normalise the amplitudes shown in Fig.\,\ref{fig:quad_comp}.

\begin{figure*}
\includegraphics[width=\textwidth]{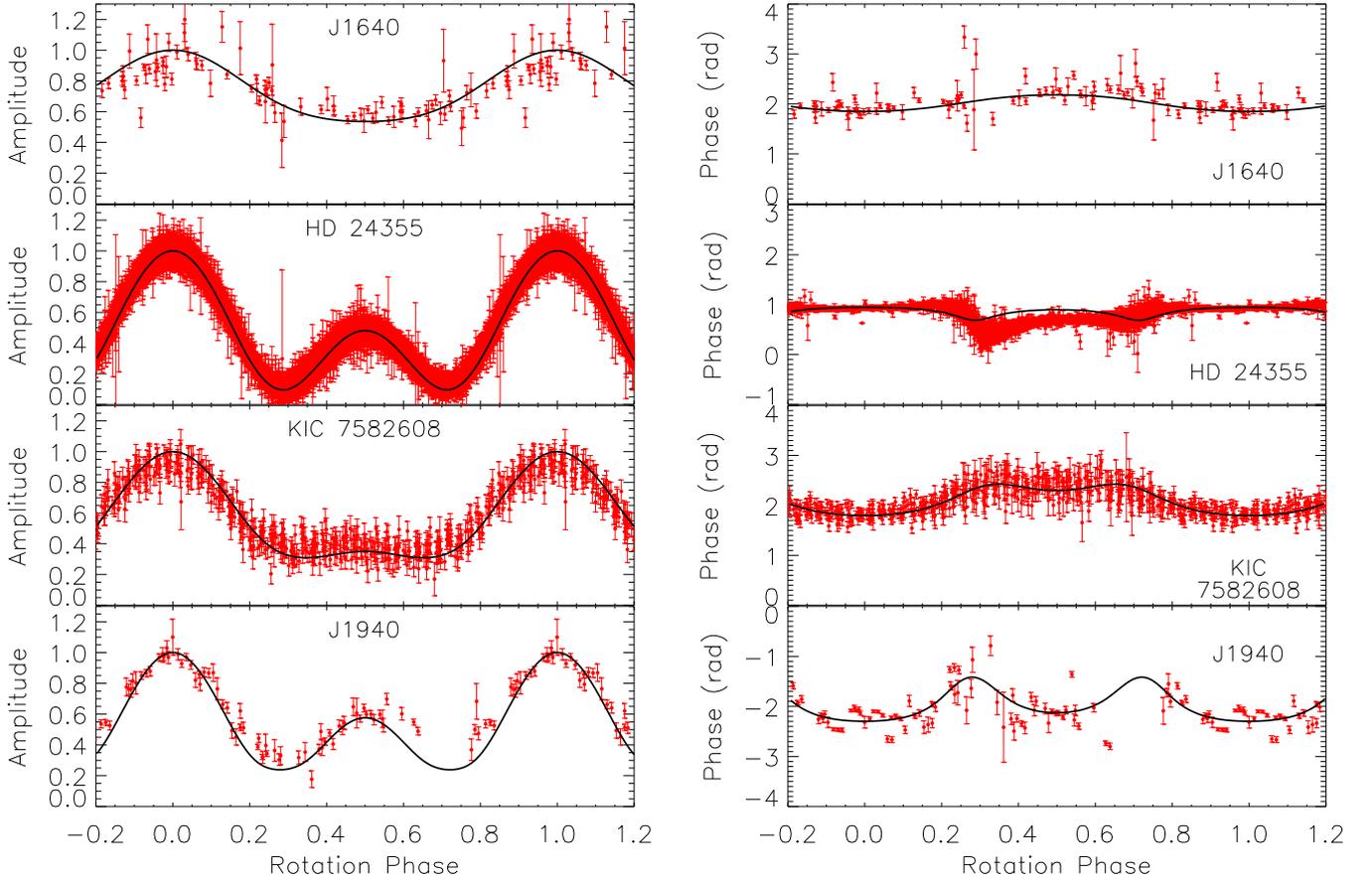}
\caption{Comparison of amplitude and phase modulations of four quadrupole roAp stars, which pulsate in distorted modes. The stars are ordered, from top to bottom, in the increasing contribution of the higher order $\ell$ components of the pulsation. The strength of the phase change becomes more apparent from top to bottom. The data are plotted in red points, with the black line representing the model fits. The amplitudes are normalised to each star's maximum amplitude.}
\label{fig:quad_comp}
\end{figure*}

\begin{table*}
\centering
  \caption{Comparison of the model parameters used to calculate the fits to the phases and amplitudes of the four stars shown in Fig.\,\ref{fig:quad_comp}; the values of $i$ and $\beta$ are interchangeable with each other. The final two columns are the measured parameters. The quoted pulsation amplitude is the peak-to-peak amplitude at pulsation maximum. The calculation of the $B$ amplitude for KIC\,7582608 and HD\,24355 is discussed in the text.}
  \label{tab:model_all}
  \begin{tabular}{lccccccccc}
    \hline
    Star	&	Mass	&	$\log g$	&	$T_{\rm eff}$	&	$B_{\rm p}$	&	$\log L/{\rm L_\odot}$	&	$i$ 		& $\beta$ 		&Pulsation Frequency	&	Pulsation Amplitude ($B$)\\
    		&	(M$_\odot$)&	(cm\,s$^{-2}$)&	(K)			&	(kG)		&							&	$(^\circ)$	& $(^\circ)$	&	(\cd)				&	(mmag)\\
    \hline
    J1640 	&	2.00		&	3.940	&	7700			&	1.70			&	1.298				&	70		&	13		& 151.93463		&	20.68  \\
    HD\,24355	& 2.00	& 	4.051	&	8100			&	1.40			&	1.279				&	45		&	77		& 224.30430		&	17.73	\\
    KIC\,7582608 & 2.10	&	4.087	&	8570			&	4.00			&	1.357				&	75		&	25		& 181.73739		&	26.40 \\
    J1940 	&	2.00		&	3.914	&	7600			&	1.55			&	1.301				&	30		&	84		& 176.38312		&	34.30  \\
    \hline
    \end{tabular}
    \end{table*}

As can be seen, there is a gradual change in the strength of the phase variations in these stars, from top to bottom. When comparing the contribution of the different spherical harmonic components of the fits, we find that more energy is being distributed to higher order terms. Most of the energy in J1640 is confined to the $\ell=0$ and $\ell=2$ components. This is almost the case for KIC\,7582608 and HD\,24355 too, however to reproduce the phase variations well, higher $\ell$ terms are required. Finally, the energy of the $\ell=0$ and $\ell=2$ components for J1940 is not sufficient to describe the phase variation, and we find significant energy is contributed from the higher order terms. We show the relative kinetic energy of each of the $\ell=0,2,4,6$ components of each of the stars in Table\,\ref{tab:k.e.}

\begin{table}
\centering
  \caption{The relative kinetic energy of the $\ell=0,2,4,6$ components for the four stars presented in Fig\,\ref{fig:quad_comp}, given the model parameters in Table\,\ref{tab:model_all}.}
  \label{tab:k.e.}
  \begin{tabular}{lcccc}
    \hline
    Star	&	\multicolumn{4}{c}{Relative Kinetic Energy} \\
    		&	$\ell=0$ & $\ell=2$ & $\ell=4$ & $\ell=6$ \\
    \hline
    J1640 		&	     0.930	&	1.000	&	0.102		&	0.017 \\
    HD\,24355	&   0.723	&	1.000	&	0.109		&	0.040 \\
    KIC\,7582608 & 1.000	&	0.795	&	0.089		&	0.072 \\
    J1940 		&	     0.619	&	1.000	&	0.391		&	0.220 \\
    \hline
    \end{tabular}
    \end{table}

There is a relation then between strength of the $\ell=0$ and $\ell=2$ components and the energy supplied by higher order terms, and the strength of the phase variation at rotation phase $0.5$. We note here that the strength of the amplitude variations is much less dependant on the components of the spherical harmonic fit, but rather by the values of $i$ and $\beta$ i.e. changing the viewing aspect of the mode. 

It is interesting to note that all four of these stars are extremely high amplitude pulsators for roAp stars. Whether, then, these distorted quadrupole modes are only present in the highest amplitude roAp stars, or whether we are being subjected to a detection bias, is yet to be seen.

Further more, when comparing the characteristics of these four stars, we find they all show (within the limit of our observations) single pulsation modes which are above the theoretical acoustic cutoff frequency with pulsation phase changes much smaller than predicted. To put some of these characteristics into perspective, we plot these stars (in red) in the $T_{\rm eff} - \nu L/M$ plane, after \citet{saio14}, in Fig.\,\ref{fig:teff_nulm} alongside other roAp stars. It is obvious from the plot that these four stars occupy a region of this plane removed from all but one of the other roAp stars; due to the low amplitude signal in the pulsation spectrum of HD\,42659 \citep{martinez94}, it has yet to be studied in detail to determine if it is also a distorted quadrupole pulsator. 

\begin{figure}
\includegraphics[width=\columnwidth]{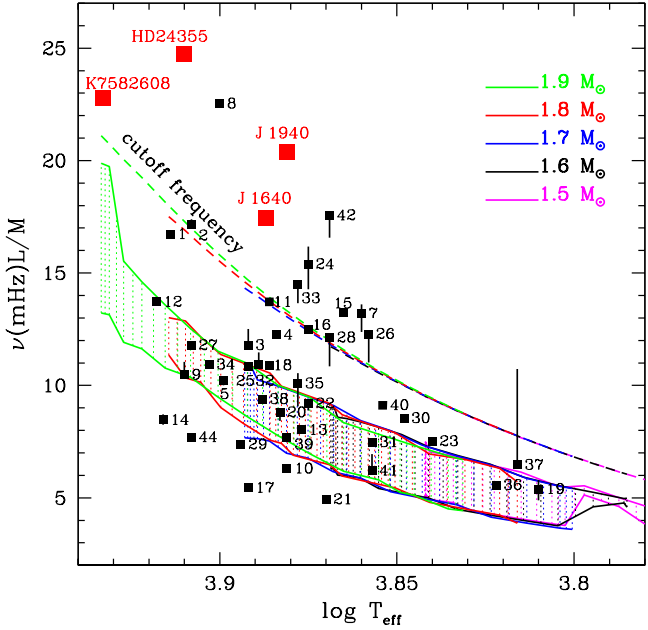}
\caption{The positions of roAp stars in the $T_{\rm eff} - \nu L/M$ plane, in which L/M is in solar units. The principal frequencies are represented by squares, with vertical bars showing the range of frequencies for multi-periodic stars. The hatched region is where the $\kappa$-mechanism excites high-order p-modes in the H-ionization zone in non-magnetic models. Acoustic cut-off frequencies are represented by the dashed lines. The red squares of J1640, J1940, HD\,24355 and KIC\,7582608 are stars that have been studied in detail and show similar characteristics which differ from the other stars. The numerical labels correspond to the stars in Table\,\ref{tab:teff_nulm_tab}. Figure after \citet{saio14}.}
\label{fig:teff_nulm}
\end{figure}

To construct Fig.\,\ref{fig:teff_nulm}, we used the effective temperatures presented in table 1 of \citet{smalley15}, and, where available, \emph{Gaia} parallaxes (from DR1) to estimate stellar luminosity \citep{gaia16,gaiaDR1}. In the cases where \emph{Gaia} parallaxes are not available, we use the $\log L/{\rm L_{\odot}}$ given in \citet{smalley15}.  Where required, the bolometric scale of \citet{flower96} was used. To determine the stellar mass, we compare the star's position with evolutionary tracks on the HR diagram. The tracks were produced with the {\sc{mesa}} code \citep{paxton13} with initial mass fractions of hydrogen and helium of $X=0.70$ and $Y=0.28$, respectively. The full parameters used for the plot are given in Table\,\ref{tab:teff_nulm_tab}. Not all roAp stars have this information available; that is why not all 61 roAp stars are plotted. The mass, luminosity and principal frequency parameters are combined to provide the $\nu L/M$ parameter; this eliminates the mass dependence of the cutoff frequency and instability range due to H-ionisation zone at a given effective temperature.

The positions, in Fig.\,\ref{fig:teff_nulm}, of the four stars discussed here have been determined with the model fits, i.e. using the parameters in Table\,\ref{tab:model_all}. However, we note that there is a \emph{Gaia} parallax (from DR1) of HD\,24355 which places the star below the zero-age main-sequence, which is contradictory to the values derived from spectroscopy and pulsation modelling \citep{holdsworth16}. We therefore adopt the model luminosity, and await a second data release for more accurate measurements of the parallax.

This discrepancy between the \emph{Gaia} derived luminosity and that from spectroscopy and pulsation modelling may be present in some of the other black points in Fig.\,\ref{fig:teff_nulm}. We do not yet have a uniform data set to construct the diagram, and so caution must be exercised when drawing conclusions. Further releases of \emph{Gaia} data, especially of the four red points, will provide better constraints on the luminosity. Furthermore, light curves obtained by the {\emph{Transiting Exoplanet Survey Satellite (TESS)}} mission \citep{ricker15} will provide us ample data to model the pulsations in the roAp stars. These two data sets will provide a nearly homogeneous data set to work with, thus removing many of the biases currently present in the data used to plot Fig.\,\ref{fig:teff_nulm}.

The identification and study of J1640 as only the fourth member of these strongly distorted pulsators is allowing us to test the extremes of the pulsations seen in the roAp stars. Previous studies of `well behaved', non-distorted, quadrupole pulsators have not allowed for the testing of the pulsation models, in the context of the contribution of high order $\ell$ components, to the extent we are able to do here.

The seemingly related subgroup of roAp stars is allowing us to test the pulsation theory which has, so far, well represented the pulsations in most of the members of this class. Although we are able to force our current models to fit these extreme members of the roAp stars, perhaps their supercritical, phase-suppressed pulsations require an entirely different pulsation driving mechanism. This may come in the form of turbulent pressure driving the pulsation, as suggested by \citet{cunha13}, but further examples and comparisons are needed to confirm this.


\section{Summary and Conclusions}

The identification of J1640 as a roAp star through a search of the SuperWASP photometric archive provided an indication that it is one of the highest amplitude roAp stars known to date, when considering the change in measured pulsation amplitude as a function of wavelength \citep{medupe98}. We have presented an analysis of the WASP data in more detail than presented in the discovery paper \citep{holdsworth14a}. Furthermore, we have presented the analysis of multisite photometric data targeting J1640 made with $B$ observations. The results show that, as expected, J1640 is one of the largest amplitude roAp pulsators.

J1640 was observed as a secondary target in the 2016 observing season at the South African Astronomical Observatory. It was clear from these observations that J1640 warranted its own dedicated follow-up. To that end, we observed the star in the 2017 season for a total of 199\,h with the 1.0-m telescope at SAAO and the Las Cumbres Observatory 1.0-m telescope network located at both the Siding Springs Observatory in Australia, and the Cerro Tololo Interamerican Observatory in Chile. The longitudinal separation of the telescopes allowed for almost continuous observations (for a limited time).

The best measure of the rotation period of J1640 comes from the analysis of the follow-up observations with a $B$ filter. The increased amplitude of the rotation signature over the WASP data allowed us to more precisely determine the period, despite the shorter time base of observations. Assuming that the spot(s) which cause the mean light variations are stable, and assuming the oblique rotator model of \citet{stibbs50}, we find a period of $3.6747\pm0.0005$\,d. This is also in agreement with the period found from the WASP data.

Analysis of the final light curve revealed that J1640 pulsates with a single mode with a central frequency, surrounded by four sidelobes split by one and two times the rotation frequency. Such a manifestation is a result of oblique pulsation of a quadrupole mode. The outermost sidelobes, however, have low amplitude, and the two inner most sidelobes are out of phase with the other peaks, leading to the conclusion that the quadrupole mode in this star is distorted.

The full light curve provides us with almost complete rotation-phase coverage of the star. As a result, we have been able to analyse the pulsation amplitude and phase as the star rotates. We find that the amplitude varies in a sinusoidal way, but the phase is almost constant -- contrary to what would be expected for a non-distorted mode. To investigate this phenomenon further, we model J1640 using the method of \citet{saio05}. The modelling results confirm that J1640 is a quadrupole pulsator, with a magnetic field of strength $1.70$\,kG distorting the mode.

Furthermore, when considering the components of the spherical harmonic series which describes the pulsation in J1640, we find that the mode is dominated by the $\ell=0$ and $\ell=2$ components, with small contributions from the higher $\ell$ terms.

The amplitude and phase variations in J1640 are reminiscent of those seen in KIC\,7582608 \citep{holdsworth14b}, HD\,24355 \citep{holdsworth16} and J1940 \citep{holdsworth18a}. They too are quadrupole pulsators with suppressed phase variations. In modelling those stars in the same way as we have presented here, we find there is a dependence on the strength of the high order $\ell$ components of the pulsation on the strength of the phase variability. The greater the phase variation, the stronger the $\ell>2$ components are. These stars all pulsate with very high amplitudes, when compared to the roAp class of stars as whole. This leads us to question whether the extreme distortions are linked to the amplitude of pulsation, or whether we are currently only able to detect this phenomenon due to observational limits.

We will be afforded the opportunity to conduct a homogeneous study of (most of) the roAp stars with the launch of the {\emph{TESS}} mission. At the expected precision of {\emph{TESS}}, we will be able to identify {\emph{all}} the distorted quadrupole (and dipole) pulsators, allowing us to test the significance of the spherical harmonic components of a large sample of stars, and to determine if the pulsation amplitude plays a significant role in the distorted stars. Until then, we will continue to study these stars with ground-based observations.
 
\section*{Acknowledgements}

DLH acknowledges financial support from the STFC via grant ST/M000877/1. The research leading to these results has received funding from the European Research Council (ERC) under the European Union's Horizon 2020 research and innovation programme (grant agreement N$^{\rm o}$\,670519: MAMSIE). MJ acknowledges financial support from the National Science Foundation via grant AST-1211384.
This paper uses observations made at the South African Astronomical Observatory (SAAO), and from the LCO 1.0-m telescope network. Some of the observations reported in this paper were obtained with the Southern African Large Telescope (SALT) under programme 2012-2-UKSC-001, and with the WHT operated on the island of La Palma by the Isaac Newton Group in the Spanish Observatorio del Roque de los Muchachos of the Instituto de Astrof\'i­sica de Canarias, in service mode. The WASP project is funded and maintained by Queen's University Belfast, the Universities of Keele, St. Andrews and Leicester, the Open University, the Isaac Newton Group, the Instituto de Astrofisica Canarias, the South African Astronomical Observatory and by the STFC. This work has made use of data from the European Space Agency (ESA) mission {\emph Gaia} (\url{https://www.cosmos.esa.int/gaia}), processed by the {\emph Gaia} Data Processing and Analysis Consortium (DPAC, \url{https://www.cosmos.esa.int/web/gaia/dpac/consortium}). Funding for the DPAC has been provided by national institutions, in particular the institutions participating in the {\emph Gaia} Multilateral Agreement.
We thank D. Bollen for obtaining a spectrum of KIC10483436 with the HERMES spectrograph, installed at the Mercator Telescope, operated on the island of La Palma by the Flemish Community, at the Spanish Observatorio del Roque de los Muchachos of the Instituto de Astrof\'isica de Canarias and supported by the Fund for Scientific Research of Flanders (FWO), Belgium, the Research Council of KU Leuven, Belgium, the Fonds National de la Recherche Scientific (F.R.S.--FNRS), Belgium, the Royal Observatory of Belgium, the Observatoire de Gen\`eve, Switzerland and the Th\"uringer Landessternwarte Tautenburg, Germany.
We thank the referee for useful comments and suggestions.

\bibliography{J1640-refs}

\appendix
\section{Parameters for $T_{\rm eff} - \nu L/M$ diagram}

\begin{landscape}
\begin{table}
\centering
  \caption{Values used to produce Fig.\,\ref{fig:teff_nulm}. Effective temperatures are from \citet{smalley15} unless stated otherwise. Luminosities are calculated from \emph{Gaia} parallaxes with the bolometric corrections from \citet{flower96}, unless stated otherwise. Column two refers to the star labels in Fig.\,\ref{fig:teff_nulm}.}
  \label{tab:teff_nulm_tab}
  \begin{tabular}{llccccccrcccl}
    \hline
    Star & Star &Main Frequency	& Min. Frequency & Max. Frequency	& Mass  	       & $\log L$ 	       &  $\log T_{\rm eff}$ 	& \multicolumn{1}{c}{Parallax} & $V$    & $M_V$ & B.C.   & Notes \\
    	   & Number& (mHz)			& (mHz)		    & (mHz)		& (M$_\odot$) & (L$_{\odot}$) & (K)				& \multicolumn{1}{c}{(mas)}     & (mag) & (mag)   & (mag) &	\\
    \hline

HD6532 		&1 & 2.402 & 2.402 & 2.402 & 1.85 & 1.11 & 3.914 & $5.131\pm0.300$ 	& 8.40 	& 1.95	& 0.019 	& \\
HD9289 		&2 & 1.585 & 1.572 & 1.605 & 2.02 & 1.34 & 3.908 & $1.256\pm0.311$ 	& 9.38 	& -- 		& -- 		& $\log L$ from \citet{smalley15} \\
HD12098 		&3 & 2.174 & 2.164 & 2.306 & 1.72 & 0.97 & 3.892 & $7.014\pm0.314$ 	& 8.07 	& 2.30 	& 0.031 	& \\
HD12932 		&4 & 1.436 & 1.436 & 1.436 & 1.90 & 1.21 & 3.884 & \multicolumn{1}{c}{--} &10.36	& -- 		& -- 		& $\log L$ from \citet{smalley15} \\
HD19918 		&5 & 1.510 & 1.481 & 1.510 & 1.82 & 1.09 & 3.899 & $3.376\pm0.225$ 	& 9.35 	& 1.99 	& 0.028 	& \\
HD24355		& & 2.596 & 2.596 & 2.596 & 2.00 & 1.28 & 3.910 & $3.994\pm0.270$ 	& 9.65 	& 2.66 	& 0.022 	& Best model from \citet{holdsworth16} \\
HD24712 		&7 & 2.721 & 2.553 & 2.806 & 1.60 & 0.89 & 3.860 & $19.808\pm0.864$ & 6.00 	& 2.48 	& 0.035 	& \\
HD42659 		&8 & 1.718 & 1.718 & 1.718 & 2.10 & 1.44 & 3.900 & $7.477\pm0.415$ 	& 6.75 	& 1.12 	& 0.028 	& \\
HD60435 		&9 & 1.381 & 1.381 & 1.457 & 1.82 & 1.14 & 3.910 & $3.944\pm0.239$ 	& 8.89 	& 1.87 	& 0.023 	& \\
HD69013 		&10 & 1.485 & 1.485 & 1.485 & 1.63 & 0.84 & 3.881 & $4.064\pm0.598$ 	& 9.56 	& 2.60 	& 0.034 	& \\
HD75445 		&11& 1.850 & 1.850 & 1.850 & 1.82 & 1.13 & 3.886 & $9.040\pm0.278$ 	& 7.12 	& 1.90 	& 0.033 	& \\
HD80316 		&12 & 2.252 & 2.252 & 2.252 & 1.80 & 1.04 & 3.918 & $7.486\pm0.299$ 	& 7.77 	& 2.14 	& 0.015 	& \\
HR3831 		&13& 1.428 & 1.428 & 1.428 & 1.70 & 0.98 & 3.877 & $16.144\pm0.762$ & 6.23 	& 2.27 	& 0.034 	& \\
HD84041 		&14& 1.113 & 1.085 & 1.145 & 1.90 & 1.16 & 3.916 & $3.121\pm0.283$ 	& 9.36 	& 1.83 	& 0.019 	& \\
HD86181 		&15& 2.688 & 2.688 & 2.688 & 1.65 & 0.91 & 3.865 & $4.072\pm0.248$ 	& 9.38 	& 2.43 	& 0.035 	& \\
HD92499 		&16& 1.602 & 1.602 & 1.602 & 1.73 & 1.13 & 3.875 & $3.996\pm0.302$ 	& 8.88 	& 1.89 	& 0.035 	& \\
HD96237 		&17& 1.200 & 1.200 & 1.200 & 1.67 & 0.88 & 3.892 & $2.800\pm0.283$ 	& 9.45 	& 1.69 	& 0.031 	& $\log L$ from \citet{smalley15} \\
HD99563 		&18& 1.558 & 1.554 & 1.562 & 1.80 & 1.10 & 3.886 & $4.457\pm0.293$ 	& 8.72 	& 1.97 	& 0.037 	& \\
HD101065 	&19& 1.373 & 1.245 & 1.474 & 1.50 & 0.77 & 3.810 & $9.110\pm0.246$ 	& 8.03 	& 2.83 	& 0.003 	& \\
HD115226 	&20& 1.373 & 1.315 & 1.382 & 1.75 & 1.05 & 3.883 & $5.201\pm0.293$ 	& 8.50 	& 2.08 	& 0.033 	& \\
HD116114 	&21& 0.782 & 0.782 & 0.782 & 1.70 & 1.03 & 3.870 & $10.595\pm0.917$ & 7.02 	& 2.14 	& 0.035 	& \\
HD119027 	&22& 1.954 & 1.888 & 1.954 & 1.65 & 0.89 & 3.875 & $3.269\pm0.463$ 	& 9.92 	& 2.49 	& 0.035 	& \\
HD122970 	&23& 1.501 & 1.501 & 1.501 & 1.55 & 0.89 & 3.840 & $6.942\pm0.342$ 	& 8.29 	& 2.50 	& 0.028 	& \\
HD128898 	&24& 2.442 & 2.265 & 2.567 & 1.70 & 1.03 & 3.875 & \multicolumn{1}{c}{--}& 3.19 	&  --		&  	--	& $\log L$ from \citet{smalley15} \\
HD132205 	&25& 2.334 & 2.334 & 2.334 & 1.67 & 0.89 & 3.892 & $5.693\pm0.284$ 	& 8.72 	& 2.50 	& 0.032 	& \\
HD134214 	&26& 2.950 & 2.647 & 2.983 & 1.55 & 0.81 & 3.858 & $11.111\pm0.680$ & 7.46 	& 2.69 	& 0.034 	& \\
HD137909 	&27& 1.031 & 1.031 & 1.031 & 2.05 & 1.37 & 3.908 & \multicolumn{1}{c}{--}& 3.68 	&--  		&  --		& $\log L$ from \citet{smalley15} \\
HD137949 	&28& 2.015 & 1.803 & 2.015 & 1.70 & 1.01 & 3.869 & $12.663\pm0.647$ & 6.69 	& 2.20 	& 0.035 	& \\
HD148593 	&29& 1.560 & 1.560 & 1.560 & 1.68 & 0.90 & 3.894 & $4.603\pm0.311$ 	& 9.15 	& 2.47 	& 0.031 	& \\
HD151860 	&30& 1.355 & 1.355 & 1.355 & 1.70 & 1.03 & 3.848 & $4.247\pm0.240$ 	& 9.01 	& 2.15 	& 0.032 	& \\
HD154708 	&31& 2.088 & 2.088 & 2.088 & 1.50 & 0.73 & 3.857 & $6.745\pm0.259$ 	& 8.76 	& 2.90 	& 0.034 	& $\log T_{\rm eff}$ from \citet{joshi16}\\
HD166473 	&32& 1.833 & 1.833 & 1.928 & 1.72 & 1.01 & 3.889 & $7.189\pm0.288$ 	& 7.92 	& 2.20 	& 0.032 	& \\
HD176232 	&33& 1.448 & 1.366 & 1.469 & 1.95 & 1.29 & 3.878 & $13.260\pm0.485$ & 5.89 	& 1.50 	& 0.034 	& $\log T_{\rm eff}$ from \citet{nesvacil13}\\
HD177765 	&34& 0.706 & 0.706 & 0.706 & 2.24 & 1.54 & 3.903 & $2.323\pm0.311$ 	& 9.15 	& 0.98 	& 0.033 	& $\log L$ from \citet{smalley15} \\
HD201601 	&35& 1.365 & 1.228 & 1.427 & 1.70 & 1.10 & 3.878 & \multicolumn{1}{c}{--}& 4.68 	&  --		& -- 		& $\log L$ from \citet{smalley15} \\
HD213637 	&36& 1.452 & 1.411 & 1.452 & 1.50 & 0.76 & 3.822 & $4.502\pm0.301$ 	& 9.58 	& 2.85 	& 0.015 	& \\
HD217522 	&37& 1.215 & 1.120 & 2.017 & 1.60 & 0.93 & 3.816 & $9.500\pm0.354$ 	& 7.52 	& 2.41 	& 0.009 	& \\
HD218495 	&38& 2.240 & 2.240 & 2.240 & 1.65 & 0.84 & 3.888 & $4.415\pm0.247$ 	& 9.38 	& 2.60 	& 0.033 	& \\
HD218994 	&39& 1.170 & 1.170 & 1.170 & 1.75 & 1.06 & 3.881 & \multicolumn{1}{c}{--}& 8.56 	&  --		&  --		& $\log L$ from \citet{smalley15} \\
HD185256 	&40& 1.613 & 1.613 & 1.613 & 1.65 & 0.97 & 3.854 & $2.912\pm0.834$ 	& 9.96 	& 2.28 	& 0.036 	& $\log L$ from \citet{kochukhov13}\\
K10483436	&41& 1.353 & 1.353 & 1.512 & 1.50 & 0.84 & 3.857 & $0.761\pm0.775$ 	& 11.38 	& -- 		&  --		& $T_{\rm eff}$ estimated from private spectrum\\
K10195926 	&42& 0.975 & 0.920 & 0.975 & 2.26 & 1.61 & 3.869 & $1.577\pm0.504$ 	& 10.66 	& 1.65 	& 0.035 	& $\log L$ from \citet{smalley15} \\
K7582608		& &2.103 & 2.103 & 2.103 & 2.10 & 1.36 & 3.933 & \multicolumn{1}{c}{--}& 11.25 & --		& --		& Model parameters derived in this work\\
K4768731 	&44& 0.711 & 0.711 & 0.711 & 2.02 & 1.34 & 3.908 &  \multicolumn{1}{c}{--}&  9.16 	&  --		&  --		& $\log L$ from \citet{smalley15} \\
J1640		& &1.758 & 1.758 & 1.758 & 2.00 & 1.30 & 3.887 & \multicolumn{1}{c}{--}& 12.70	&	--	& --		& This work\\
J1940		& &2.041 & 2.041 & 2.041 &2.00 & 1.30 & 3.881 & \multicolumn{1}{c}{--}& 13.10	&	--	&--		& Best model from \citet{holdsworth18a}\\
    
    \hline
    \end{tabular}
    \end{table}
\end{landscape}

\label{lastpage}
\end{document}